\documentclass[manuscript]{aastex}  
\usepackage{epstopdf}
\usepackage{ulem}
\usepackage{amssymb}
\usepackage{natbib}
\usepackage{times}
\usepackage{graphicx}
\usepackage[usenames,dvipsnames]{pstricks}
 \usepackage{psfrag}

\def\aj{AJ}%% Astronomical Journal
\def\araa{ARA\&A}%% Annual Review of Astron and Astrophys
\def\apj{ApJ}%% Astrophysical Journal
\def\apjl{ApJ}%% Astrophysical Journal, Letters
\def\apjs{ApJS}%% Astrophysical Journal, Supplement
\def\ao{Appl.~Opt.}%% Applied Optics
\def\apss{Ap\&SS}%% Astrophysics and Space Science
\def\aap{A\&A}%% Astronomy and Astrophysics
\def\aapr{A\&A~Rev.}%% Astronomy and Astrophysics Reviews
\def\mnras{MNRAS}%% Monthly Notices of the RAS
\def\prd{Phys.~Rev.~D}%% Physical Review D
\def\pasj{PASJ}%% Publications of the ASJ
\def\physrep{Phys.~Rep.}%% Physics Reports
\newcommand{\be}{\begin{equation}}
\newcommand{\ee}{\end{equation}}
\newcommand{\bary}{\begin{eqnarray}}
\newcommand{\eary}{\end{eqnarray}}
\newcommand{\en}{E_\nu}
\shorttitle{ Centaurus A?}
\shortauthors{Fraija N.}

\begin{document}
\title{Correlation of $\gamma$-ray and high-energy cosmic ray  fluxes from the giant lobes of Centaurus A}
\author{N. Fraija\altaffilmark{1}}
\affil{Instituto de Astronom\'ia, UNAM, M\'exico, 04510}
\email{nifraija@astro.unam.mx}
\altaffiltext{1}{Luc Binette-Fundaci\'on UNAM Fellow. Instituto de Astronom\' ia, Universidad Nacional Aut\'onoma de M\'exico, Circuito Exterior, C.U., A. Postal 70-264, 04510 M\'exico D.F., M\'exico}

\date{\today} 
	
\begin{abstract}
The spectral energy distribution of giant lobes shows one main peak detected by the Wilkinson Microwave Anisotropy Probe at low energy of  $10^{-5}$ eV  and  a faint $\gamma$-ray  flux  imaged by Fermi Large Area Telescope at  energy $\geq$ 100 MeV.  On the other hand, Pierre Auger Observatory associated some ultra-high-energy cosmic rays with the direction of Centaurus A  and IceCube reported 28 neutrino-induced events in a TeV - PeV energy range,  although none of them related with this direction. In this work  we describe the spectra for each of the lobes, the  main peak with synchrotron  radiation, and the high-energy emission with  $pp$ interactions.  Obtaining a good description of the main peak, we deduce  the magnetic fields, electron densities and the age of the lobes. Describing successfully the $\gamma$-ray emission by  pp interactions and considering as targets those  thermal particles in the lobes with  density in the range $10^{-10}$ to  $10^{-4}\, {\rm cm}^{-3}$,  we calculate the number of ultra-high-energy cosmic rays. Although  $\gamma$-spectrum is well described with any density in the range, only when  $10^{-4}\, {\rm cm}^{-3}$  is considered, the expected number of events is very similar to that observed by  Pierre Auger Observatory, otherwise we obtain an excessive luminosity.  In addition,  correlating  the $\gamma$-ray and neutrino fluxes through pp interactions we calculate the number of  high-energy neutrinos expected  in IceCube.  Our analysis indicates that neutrinos above 1 TeV  cannot be produced in the  lobes of Centaurus A, which is consistent with the results recently published by IceCube Collaboration.
%PACS numbers may be entered using the \verb+\pacs{#1}+ command.
\end{abstract}


%\pacs{98.70.Rz; 98.70.Sa}% PACS, the Physics and Astronomy
%\maketitle

\section{Introduction}
Centaurus A (Cen A), at a distance of 3.8 Mpc, is the nearest radio-loud active galactic nucleus (AGN).  Due to its distance, Cen A is an excellent source for studying the physics of relativistic outflows and radio lobes.  It has a  jet with an axis subtending an angle to the line of sight estimated as $15^\circ\,-\,80^\circ$ \citep[see, e.g.] [and reference therein]{2006PASJ...58..211H} and two  giant lobes oriented primarily in the north-south direction, which subtend $\sim\,10^\circ$ on the sky. They were imaged and analyzed by Parkes radio telescope at 6.3 cm \citep{1993A&A...269...29J, 2000A&A...355..863A} and at 22, 33, 41, 61 and 94  GHz by the Wilkinson Microwave Anisotropy Probe  \citep[WMAP;][]{2009ApJS..180..225H,2003ApJS..148...39P, 2009MNRAS.393.1041H, 2010Sci...328..725A}.    Also for a period of 10 months, Cen A was monitored by Large Area Telescope (LAT) on board the Fermi Gamma-Ray Space Telescope  \citep{2009ApJ...697.1071A}  and $\gamma$-ray excesses were detected  from both lobes.  The resulting LAT image showed the $\gamma$-ray peak coincident with the active galactic nucleus detected by the Compton/EGRET instrument \citep{1999ApJS..123...79H}.  Assuming a power law for the $\gamma$-ray spectra and from the resultant test statistics \citep{1996ApJ...461..396M} LAT recorded a flux of $[0.77(+0.23/-0.19)_{\rm stat}(\pm0.39)_{\rm syst}] \times 10^{-7} {\rm ph\,cm^{-2}\,s^{-1}}$ with photon index $2.52(+0.16/-0.19)_{\rm stat}(\pm0.25)_{\rm syst}$ for the north lobe and a flux of $[1.09(+0.24/-0.21)_{\rm stat}(\pm0.32)_{\rm syst}] \times 10^{-7} {\rm ph\,cm^{-2}\,s^{-1}}$  with photon index $2.60(+0.14/-0.15)_{\rm stat}(\pm0.20)_{\rm syst}$ for the south lobe \citep{2010Sci...328..725A}.\\
\citet{2006MNRAS.368L..15H} have claimed that the oncoming jet enters the northern inner lobe, encrusted in the thermal interstellar gas of NGC5128, at $\sim$ 3.5 kpc.  Based on deep Chandra observations,  \citet{2007ApJ...670L..81H} reclaimed  that the receding jet extends out to $\sim$ 2.5 kpc in protection in X rays, showing also up on a similar scale in radio \citep{2003ApJ...593..169H, 1998AJ....115..960T}.   Based on the detection of extended thermal X-ray emission from this region, \citet{2009ApJ...698.2036K} interpreted the northern middle lobe as an old structure that has recently become reconnected to the energy supply from the jet \citep{2013A&A...558A..19W}. \\   
 Based on X-ray (0.5-2.5 keV) measurements and supposing that all the emission comes from a uniform thermal plasma, \citet{2009MNRAS.393.1041H} established  an strict upper limit on this plasma, that is $n_p\sim 10^{-4} cm^{-3}$.  Recently,  considering the internal Faraday rotation scenario,  \citet{2013ApJ...764..162O} presented a positive detection of the internal depolarization signal leading to  the same value of density.  Also, \citet{2013ApJ...766...48S} presenting an analysis of the diffuse X-ray emission  found a tentative detection of a soft excess component with an energy of $kT\sim 0.5\, keV$, corresponding to the same value of the number density of the thermal gas.  However,  \cite{2013A&A...558A..19W}   estimating the values of total entrainment, buoyancy age and the average volume of the giant lobes found a different number density of thermal particles  $n_p\sim 10^{-9} cm^{-3}$.   In addition, they calculated  that  the relativistic electron number densities for four  giant lobe sectors defined by  \citet{2009MNRAS.393.1041H}  were in the range $ 1.0\times 10^{-11}\, {\rm cm^{-3}}   \leq N_e \leq 1.5\times 10^{-8}\,{\rm cm^{-3}} $.\\
Otherwise, based on the report given by  Pierre Auger Collaboration (PAO) with respect to the anisotropy in the arrival direction of ultra-high-energy cosmic rays (UHECRs)   \citep{2007Sci...318..938P, 2008APh....29..188P} and the possible correlation with Cen A,  some authors have pointed out that Cen A has the potential to accelerate protons up to ultra-high energies \cite[e.g.][]{2008JETPL..87..461G,2009ApJ...693.1261M, 2009NJPh...11f5016D}. \\
Recently, IceCube reported the detection of events in an  energy range of TeV - PeV \citep{2013arXiv1311.5238I, 2013arXiv1304.5356I} and although these  events have been discussed to have an extragalactic origin\citep{2012arXiv1211.1974C,  2013ApJ...766...73L, PhysRevLett.111.121102, 2013arXiv1307.7596R, 2013MNRAS.tmp.2798F}, they were not correlated with the direction of Cen A.\\
On the other hand, although  energy ranges in radio, infrared, optical \citep{1975ApJ...199L.139W, 1976ApJ...206L..45M, 1970ApJ...161L...1B,  1981ApJ...244..429B},  X-ray and $\gamma$-rays (MeV-TeV) \citep{2010ApJ...719.1433A,1999APh....11..221S,2009ApJ...695L..40A} have been detected close to the core of Cen A,  only photons in radio and $\gamma$-rays  have been collected from the lobes and,  while the spectral energy distribution (SED) of each lobe has been described with leptonic models; Radio (WMAP) data through synchrotron radiation and  Fermi-LAT data through  inverse Compton-scattered (IC) radiation from the cosmic microwave background (CMB) \citep{ 1986A&A...164L..16C, 1970RvMP...42..237B}   and extragalactic background light  (EBL)\citep{2010Sci...328..725A, 2009MNRAS.393.1041H, 2012AIPC.1505..590Y,  2009MNRAS.393.1041H,  2001ARA&A..39..249H,  2008ApJ...686L...5G, 2008IJMPD..17.1515R}, the SED near the core has been successfully described through  synchrotron self-Compton (SSC), and leptonic and hadronic models; SSC to fit the two main peaks \citep{2010ApJ...719.1433A, 2001MNRAS.324L..33C,2008A&A...478..111L} and  leptonic \citep{2011MNRAS.415..133H} and hadronic models to explain the  flux at TeV energies  \citep{2009NJPh...11f5016D, 2012ApJ...753...40F, 2012PhRvD..85d3012S}. In addition, these authors extrapolating the hadronic model (pp or p$\gamma$ interactions) to energies of $\sim10^{20}$ eV, have correlated the number of UHECRs observed by PAO with the flux in TeV energy range. Also, based on these observations,  \citet{2008PhRvD..78b3007C} and \citet{2008arXiv0802.0887H} have forecasted the expected  rate of HE neutrinos in IceCube.\\
In this work we develop a leptonic and hadronic model to describe the SED for each of the lobes.  For the leptonic model,  we present the synchrotron emission to fit WMAP data and for the hadronic model, we introduce pp interactions to fit Fermi-LAT data. Also other  mechanisms of emission (IC scattering  of CMB and EBL as well as p$\gamma$ interactions)  are discussed. Correlating the $\gamma$-ray, UHECR and neutrino fluxes through pp interactions, we  extend the proton and neutrino spectra through a simple and broken power law up to energy ranges of PAO and IceCube, respectively  to estimate the number of expected events in each of the experiments.
\section{Emission Processes from Lobes}
Detections of non-thermal radiation and soft X-rays from lobes have pointed out that they could be filled with a magnetized plasma and thermal particles  \citep{2009MNRAS.393.1041H, 2013ApJ...764..162O, 2013A&A...558A..19W, 2013ApJ...766...48S,2010Sci...328..725A}. Based on this approach  we are going to develop synchrotron radiation as leptonic process and pp interactions as hadronic process to describe the whole energy range of the lobes.
\subsection{Synchrotron Radiation}
The non-thermal radio emission  can be inferred through synchrotron radiation  generated  by an electron distribution.  The population of  these accelerated electrons  can be described by a broken power-law given by \citep{1994hea2.book.....L,  2001MNRAS.326.1499H, 2006MNRAS.368L..15H, 2011MNRAS.415..133H}
\begin{equation}
\label{espele}
N_e(\gamma_e)   = A_e
\cases {
\gamma_e^{-\alpha} 						& 	$\gamma_{e,m}<\gamma_e < \gamma_{e,b}$,\cr
\gamma_{e,b}    \gamma_e^{-(\alpha+1)}          & 	$\gamma_{e,b} \leq  \gamma_e<\gamma_{e,max}$,\cr
}
\end{equation}
\noindent where  $A_e$ is the proportionality electron constant, $\alpha$ is the spectral index and $\gamma_{e,i}$ are  the electron Lorentz  factors. The index $i$ is m, b or max for minimum, break and maximum, respectively.  Assuming an equipartition of energy density  between magnetic field  $U_B=B^2/8\pi$ and electrons $U_e=m_e \int\gamma_e N_e(\gamma_e)d\gamma_e$, the electron Lorentz factors are
\bary\label{lorentz}
\gamma_{e,m}&=&\frac{(\alpha-2)}{m_e(\alpha-1)}\,\frac{U_e}{N_e}\cr
\gamma_{e,b}&=& \frac{3\,m_e}{4\,\sigma_T\beta^2} \,  U_B^{-1}\,t^{-1}_{syn}  \cr
\gamma_{e,max}&=&\biggl(\frac{9\, q_e^2}{8\pi\, \sigma_T^2\,\beta^4}\biggr)^{1/4}\,U_B^{-1/4},
\eary 
\noindent  where  the constants  $m_p$, $m_e$, $q_e$ and $\sigma_T$ are the proton and electron mass, the electric charge and  Thomson cross section, respectively,  $\beta=v/c\sim 1$ and $z$=0.00183 is the redshift\citep{1998A&ARv...8..237I}.  The observed photon energies, ${\small \epsilon_\gamma^{obs}(\gamma_e)=\sqrt{\frac{8\pi q_e^2}{m_e^2}} (1+z)^{-1}\,\delta_D\, U_B^{1/2}\, \gamma^2_{e,i}}$,  for each Lorentz factor  (eq. \ref{lorentz}) are
\begin{eqnarray}\label{synrad}
\epsilon^{obs}_{\gamma,m} &=& \frac{\sqrt{8\pi}\,q_e(\alpha-2)^2}{m_e^3\,(\alpha-1)^2}\,(1+z)^{-1}\,\delta_D\,U_B^{1/2}\, U_e^2 N_e^{-2},\cr
\epsilon^{obs}_{\gamma,c} &=&\frac{ 9\sqrt{2\pi}\,q_e\,m_e}{8\,\sigma_T^2\beta^4} (1+z)^{-1}\,\delta_D\, U_B^{-3/2}\, t_{syn}^{-2},\cr
\epsilon^{obs}_{\gamma, max} &=&\frac{3\,q_e^2}{m_e\,\sigma_T\,\beta^2} (1+z)^{-1}\, \delta_D,
\end{eqnarray}
where we have applied the synchrotron cooling time scale,
\be
t_{syn}=\frac{E'_e}{(dE_e/dt)'}=\frac{3m_e^2}{4\sigma_T\beta^2}U_B^{-1}\,E^{'-1}_e
\ee
and $\delta_D$ is the Doppler factor.  On the other hand, the synchrotron spectrum is obtained  by the shape of the electron spectrum (eq. \ref{espele}) rather than  the emission spectrum of a single particle. Therefore, the energy radiated in the range $\epsilon_\gamma$ to $\epsilon_\gamma + d\epsilon_\gamma$ is given by electrons between   $E_e$ and $E_e + dE_e$; then we can estimate the photon spectrum through emissivity $\epsilon_\gamma N_\gamma(\epsilon_\gamma) d\epsilon_\gamma=\biggl(- \frac{dE_e}{dt}\biggr)\,N_e(E_e)dE_e$.  Following  \cite{1994hea2.book.....L} and \cite{1986rpa..book.....R}, it is easy to show that if electron distribution has  spectral indexes  $\alpha$ and $(\alpha-1)$, then the photon distribution has  spectral indexes $p=(\alpha-1)/2$ and $p=\alpha/2$, respectively. The proportionality constant is estimated  calculating the total number of radiating electrons  in the actual volume, ${\small n_e=N_e/V=4\pi N_e\,r_d^3/3}$, the maximum radiation power   ${\small P^{obs}_{\nu,max}\simeq  \frac{dE/dt}{\epsilon_\gamma(\gamma_e)}}$ and the distance D$_z$ from the source. Then,  we can obtain the observed synchrotron spectrum as follow
\begin{equation}
\label{espsyn}
\epsilon^2_\gamma N_\gamma(\epsilon_\gamma) = A_{syn,\gamma}
\cases {
(\frac{\epsilon_\gamma}{\epsilon_{\gamma,m}})^{4/3}    &  $\epsilon^{obs}_\gamma < \epsilon^{obs}_{\gamma,m}$,\cr
 (\frac{\epsilon_\gamma}{\epsilon_{\gamma,m}})^{-(\alpha-3)/2}  &  $\epsilon^{obs}_{\gamma,m} < \epsilon^{obs}_\gamma < \epsilon^{obs}_{\gamma,c}$,\cr
(\frac{\epsilon_{\gamma,c}}{\epsilon_{\gamma,m}})^{-(\alpha-3)/2}    (\frac{\epsilon_\gamma}{\epsilon_{\gamma,c}})^{-(\alpha-2)/2},           &  $\epsilon^{obs}_{\gamma,c} < \epsilon^{obs}_\gamma < \epsilon^{obs}_{\gamma,max} $\cr
}
\end{equation}
\noindent where
\be
\label{Asyn}
 A_{syn,\gamma}= \frac{P^{obs}_{\nu,max} n_e}{4\pi D_z^2}\,\epsilon^{obs}_{\gamma,m} \simeq \frac{8\pi\sigma_T\,\beta^2\,(\alpha-2)^2}{9\,m^2_e\,(\alpha-1)^2\,D_z^2}(1+z)^{-2}\,\delta^2_D\,U_B\,U^2_e\,N_e\,r_d^3
\ee
It is important to clarify that $r_d$ is the region where emitting electrons  are confined.  \noindent Eq. \ref{espsyn} represents the peak at lower energies (radio wavelength) of the  SED for each of the lobes. 
\subsection{PP interactions}
We suppose that accelerated protons are  cooled down through pp interactions  \citep{2008PhR...458..173B,2003ApJ...586...79A,2009herb.book.....D,2002MNRAS.332..215A}.   Pp interactions are given mainly through
\begin{eqnarray}
p\,+ p &\longrightarrow& \pi^++\pi^-+\pi^0 + X.
\label{pp}
\end{eqnarray}
\noindent Taking into account that  neutral pions decay  in two gammas,  $\pi^0\rightarrow \gamma\gamma$, and the minimum energy of photo-pion, $E_{\pi^0,{\rm min}}$, at rest frame is $ m_{\pi^0}$=139.57 MeV, then the minimum observed energy is
\be
\epsilon^{obs}_{\gamma,\pi^0,{\rm min}}\simeq\frac{\delta_D}{(1+z)}\,\frac{m_{\pi^0}}{2}.
\ee
Also charged pions  decay in neutrinos as follows 
\begin{eqnarray}
\pi^{+}\rightarrow\mu^{+}+\nu_{\mu}&\rightarrow&
e^{+}+\nu_{e}+\overline{\nu}_{\mu}+\nu_{\mu},\\
\pi^{-}\rightarrow\mu^{-}+\overline{\nu}_{\mu}&\rightarrow& e^{-}+\overline{\nu}_{e}+\nu_{\mu}+\overline{\nu}_{\mu}\,,
\end{eqnarray}
hence neutrino flux is expected to be accompanied by a $\gamma$-ray flux.\\
Assuming that accelerated protons interact with thermal particles whose number density lies in the range  $10^{-10}\,{\rm cm}^{-3}\leq n_p\leq 10^{-4}\,{\rm cm}^{-3}$ \citep{2009MNRAS.393.1041H, 2013ApJ...764..162O, 2013A&A...558A..19W, 2013ApJ...766...48S}, then  the  $\gamma$-ray spectrum, $(dN_\gamma/d\epsilon_\gamma)_{\pi^0}$, produced by  pp interactions  is \citep{2003ApJ...586...79A, 2012ApJ...753...40F, 2002MNRAS.332..215A, 2009MNRAS.393.1041H}
\be
f_{\pi^0 , pp}(E_p)\,E_p\,\left(\frac{dN_p}{dE_p}\right)^{obs}\,dE_p=\epsilon_{\gamma, {\pi^0}}\,\left(\frac{dN_\gamma}{d\epsilon_\gamma}\right)^{obs}_{\pi^0}\,d\epsilon_{\gamma, {\pi^0}},
\ee
here  $f_{\pi^0,pp}\approx t_{lobe}/t_{pp}= t_{lobe}\,n_p\,k_{pp}\,\sigma_{pp}$ is  the fractional power released,  $\sigma_{pp}\simeq 30(0.95 +0.06\,\rm{ln(E/GeV)}$ mb  is the nuclear interaction cross section, $k_{pp}=1/2$ is the inelasticity coefficient, $n_p$ is the thermal particle density, $t_{lobe}$  is the age of the lobe,  and $t_{pp}$ is the characteristic cooling time for this process.     Taking into account that  a pion carries 33$\%$ of the proton energy ($\xi$=0.33) and supposing that   $\gamma$-ray  spectrum  at GeV energy range is produced by a simple proton power law,
\begin{equation}
\label{esppr1}
\left(\frac{dN_p}{dE_p} \right)^{obs}  = A_p \left(\frac{E^{obs}_p}{{\rm GeV}}\right)^{-\alpha},
\end{equation}
where  $A_p$ is the proportionality constant normalized to GeV and $\alpha$ is the  spectral index,  then the observed $\gamma$-ray spectrum is 
\begin{equation}
\label{pp}
\left(\epsilon^{2}_\gamma\, \frac{dN_\gamma}{d\epsilon_\gamma}\right)^{obs}_{\pi^0}= A_{pp,\gamma}\, \left(\frac{\epsilon^{obs}_{\gamma,\pi^0}}{{\rm GeV}}\right)^{2-\alpha},
\end{equation}
where
\be
A_{pp,\gamma}= f_{\pi^0,pp}\,(2/\xi)^{2-\alpha}\,A_p \,{\rm GeV}^2
\label{App}
\ee
and the proton luminosity, {\small $L_p\simeq 4\pi F_p= 4\pi \int E_p \frac{dN_p}{dE_p}dE_p$}, can be written as 
\be\label{lum}
L_p=\frac{4\,\pi\,(\xi/2)^{2-\alpha} }{(\alpha-2)}\,D^2_z\,f_{\pi^0,pp}^{-1}\,A_{pp,\gamma}\,\biggl(\frac{E_{p,min}}{\,{\rm GeV}}\biggr)^{2-\alpha},
\ee
here $E_{p,min}$ corresponds to the proton energy at GeV energies. Eq. \ref{pp} shows the contribution of pp interactions to  the $\gamma$-ray spectrum for each of the lobes. 
\section{Production of UHE cosmic rays}
It has been proposed that astrophysical sources accelerating UHECRs could  produce HE $\gamma$-rays and neutrinos  by proton interactions with photons at the source and/or the surrounding radiation and matter.    We propose that the spectrum of accelerated protons  is extended from GeV  to $\sim 10^{20}$ eV energies and can be also determined through signature ($\gamma$-ray flux) produced at GeV energies.  This $\gamma$-ray flux is correlated with proton flux through eq. \ref{App}.  In addition,  we correlate the $\gamma$-ray and neutrino fluxes to find  the parameters of neutrino spectrum \citep{2008PhR...458..173B}.  Based on these correlations, we are going to calculate the  number of events for these spectra at energy ranges of  PAO and IceCube.
\subsection{UHE protons}
PAO studying the composition of the high-energy showers  found that the distribution of their properties  was situated in somewhere between pure p and pure Fe at 57 EeV\citep{2008ICRC....4..335Y, 2008APh....29..188P, 2007AN....328..614U}.  By contrast,  HiRes data are consistent with a dominant proton composition at these energies, but uncertainties in the shower properties  \citep{2007AN....328..614U} and in the particle physics extrapolated to this extreme energy scale \citep{2008ICRC....4..385E} preclude definitive statements about the composition. At least two events of the UHECRs observed by PAO were detected \citep{2007Sci...318..938P,2008APh....29..188P} inside a $3.1^{\circ}$ circle centered at Cen A.\\
\subsubsection{Mechanisms of UHECR acceleration} 
The maximum  energy required for acceleration of UHECRs is  limited by both the size ($R$) and magnetic field ($B$) of the emission region,  $E_{max}=Ze\,B\,R\,\Gamma$  \citep{1984ARA&A..22..425H}. Additional limitations are mainly due to radiative losses or available time when particles diffuse through the magnetized region.    In Cen A, a  short distance ($\sim 10^{15} {\rm cm}$) from the black hole (BH), the emission region is limited by the variability time scale $R=r_d=\frac{c\, \delta_D}{(1+z)^2}\, dt^{obs}$, hence the maximum energy required is \citep{2010ApJ...719.1433A,  2012PhRvD..85d3012S}
\be\label{sregion}
E_{max}=4\times 10^{19}\,{\rm eV}\, B_{0.8}\,dt^{obs}_{5}\,\Gamma_{0.85}
\ee
A hundred of kpc distance from the BH,  particles are accelerated inside the lobes, therefore the emission region is limited by the size of the lobes, then the maximum energy is 
\be
E_{max}=Ze\,B\,R\,\Gamma
\ee
with R=100 kpc  corresponding to a volume of $V=1.23\times 10^{71}\,cm^{-3}$ and $B$ the magnetic field of lobes.  The acceleration and diffuse  time scales are
\be\label{tac}
t_{acc}\simeq  2\pi\frac{E_{max}}{eB}
\ee
and
\be\label{tdif}
t_{diff}\simeq \frac{3}{2\pi}\frac{R^2eB} {E_{max}}
\ee
respectively.    As the lobes are inflated by jets in the surrounding medium, accelerated protons are injected inside by the jet, and confined within the lobes, by means of resonant Fermi-type processes \citep{2009MNRAS.393.1041H}.   The  non-thermal and the upper limit thermal  pressure in the lobes are $p_{nth}\simeq (U_e+U_B+U_p)$  and p$_{th}=n_p\,K\,T$, respectively, where $U_p\sim 2\, L_p\, t_{lobe}/V$ is the energy density of accelerated proton, k is the Boltzmann constant and T is the temperature. Assuming an equipartition between magnetic field and relativistic electron, $U_e=\lambda_{e,B}\,U_B$,  then the non-thermal pressure and the total energy can written as
\be\label{pressure}
p_{nth}\simeq U_B(1+\lambda_{e,B})+ 2\, L_p\frac{t_{lobe}}{V},
\ee
and
\be\label{Etotal}
E_{tot}\simeq \,U_B\,V(1+\lambda_{e,B})+ 2\, L_p\,t_{lobe},
\ee
respectively.   As one can observe from the values of the emission region in the jet (eq. \ref{sregion}),  protons might be or not  accelerated up to energies above 40 EeV, depending on the variability  time scale and strength of the magnetic field. Hence,  it is important to mention that protons could have a hybrid acceleration mechanism,  partially in the jet and finally in the lobes.\\
On the other hand,  supposing  that the  BH jet has the power to accelerate particles  up to ultra-high energies through Fermi processes, then  from the equipartition magnetic field $\epsilon_B$ and  during flaring intervals for which the apparent isotropic luminosity can reach $\approx 10^{45}$ erg s$^{-1}$, one can derive the maximum particle energy of accelerated UHECRs as \citep{2009NJPh...11f5016D, 2012ApJ...753...40F}
\begin{equation}
E_{max}\approx 1.0\times10^{20}\,\frac{Ze}{\phi}\frac{\sqrt{\epsilon_B\,L/10^{45}\, erg\, s^{-1}}}{\beta^{3/2}\,\Gamma}\,eV,
\end{equation}
\noindent where  $\Gamma=1/\sqrt{1-\beta^2}$, $\phi\simeq 1$ is the acceleration efficiency factor and Z is the atomic number. \\
Other more exotic mechanisms   that have been described in the literature are magnetic reconnection and stochastically acceleration by temperatures. In the magnetic reconnection framework, the free energy stored in the helical configuration can be converted to particle kinetic energy  in the region where the un-reconnected (upstream) magnetized fluid converges into the reconnection layer,   resulting in a continuously charged particle acceleration.   Some authors \citep{2000ApJ...530L..77B, 2010MNRAS.408L..46G,  2013A&A...558A..19W} have proposed that this mechanism might  accelerate  UHECRs either in the jet or in the giant lobes. Recently,   \citet{2013A&A...558A..19W} proposed that UHECRs could be stochastically accelerated by  high temperatures, being responsible for  the self-consistency between the entrainment calculations and the missing pressure in the lobes. 
\subsubsection{The expected Number of  UHECR } 
To determine the number of UHECRs,  we take into account the PAO  exposure, which  for a point source is given by $\Xi\,t_{op}\, \omega(\delta_s)/\Omega_{60}$, where $\Xi\,t_{op}=(\frac{15}{4})\,9\times10^3\,\rm km^2\,yr$, $t_{op} $ is the total operational time (from 1 January 2004 until 31 August 2007),  $\omega(\delta_s)\simeq 0.64$ is an exposure correction factor for the declination of Cen A, and $\Omega_{60}\simeq\pi$ is the Auger acceptance solid angle \citep{2008PhRvD..78b3007C}.    The expected number of UHECRs for each of the lobes of  Cen A observed above an energy of $60\,{\rm EeV}$ is given by
\be
N_{\tiny UHECR}= ({\rm  PAO\, Expos.})\times \,N_p,
\label{num}
\ee
where $N_p$ is calculated from a simple and broken power law of the proton spectrum at energies higher than 60 EeV.   In the first case, considering a simple  power law eq. (\ref{esppr1}) and from eqs. (\ref{App}) and (\ref{num}), the expected number of UHECRs  is
\bary
N_{\tiny UHECR}=\frac{\Xi\,t_{op}\, \omega(\delta_s)\, (\xi/2)^{2-\alpha}}{\Omega_{60}\,(\alpha-1)} \,f_{\pi^0,pp}^{-1}\,A_{pp,\gamma}\,\left(\frac{60 EeV}{\,{\rm GeV}}\right)^{-\alpha+2} \,{\rm EeV}^{-1}.
\label{nUHE1}
\eary
that corresponds to an isotropic UHECR luminosity  \citep{2009NJPh...11f5016D}
\be\label{lum}
L_{\tiny UHECR}=\frac{4\,\pi\,(\xi/2)^{2-\alpha} }{(\alpha-2)}\,D^2_z\,f_{\pi^0,pp}^{-1}\,A_{pp,\gamma}\,\biggl(\frac{60\,{\rm EeV}}{\,{\rm GeV}}\biggr)^{2-\alpha}
\ee
In the second case,  we assume  that  the proton spectrum is not extended continually up to $\sim 10^{20}$, but  broken at  some energy less than 60 EeV. Hence, it  can be written as
\begin{equation}
\label{esppr2}
\frac{dN_p}{dE_p}   = A_p
\cases {
(\frac{E_p}{{\rm GeV}})^{-\alpha} 						& 	$E_p < E_{p,b}$\cr
(\frac{E_{p,b}}{{\rm GeV}})^{-\alpha+\beta} (\frac{E_p}{{\rm GeV}})^{-\beta}          & 	$E_{p,b} \leq  E_p$,\cr
}
\end{equation}
 where $\beta$ and $E_{p,b}$  are the higher spectral index and break proton energy, respectively.  In this case, the number of expected events  is
\bary
N_{\tiny UHECR}=\frac{\Xi\,t_{op}\, \omega(\delta_s)\, (\xi/2)^{2-\alpha}}{\Omega_{60}\,(\alpha-1)\,} \,f_{pp}^{-1}\,A_{pp,\gamma} \,\left(\frac{E_{p,b}}{\,{\rm GeV}}\right)^{-\alpha+\beta}\,\left(\frac{60\,{\rm EeV}}{\,{\rm GeV}}\right)^{-\beta+2}\,{\rm EeV}^{-1}
\label{nUHE1}
\eary
where in eq. (\ref{nUHE1}),   $\beta$ and $E_{p,b}$ are the unknown quantities.
\subsection{HE Neutrinos}
Neutrinos are detected when they interact inside the instrumented volume. The path length L($\theta$) traversed within the detector volume by a neutrino with zenith angle $\theta$ is determined by the detector's geometry. At a first approximation, neutrinos are detected if they interact within the detector volume, i.e. within the instrumented distance  L($\theta$).  The probability of interaction  for a neutrino with energy $\en$ is 
\be
P(\en)=1-\exp\biggl[ -\frac{L}{\lambda_\nu(E_{\en})}  \biggr]\cong\frac{L}{\lambda_\nu(\en)},
\ee
where the mean free path in ice is
\be
\lambda_\nu(\en)=\frac{1}{\rho_{ice}\,N_A\,\sigma_{\nu N}(\en)}.
\ee
Here $\rho_{ice}$=0.9 g cm$^{-3}$ is the density of the ice, N$_A$=6.022$\times$ 10$^{23}$ g$^{-1}$ and $\sigma_{\nu N}$ is the neutrino-nucleon cross section.  A neutrino flux, dN$_\nu$/d$\en$, crossing a detector with energy threshold $\en^{th}$ and cross-sectional area $A(\en, \theta)$ facing the incident beam will produce
\be
N_{ev}=T\,\int_{\en^{th}} A(\en)\, P(\en)\, \frac{dN_\nu}{d\en}\,d\en,
\ee
events after a time $T$.  Furthermore, the "effective" detector area A($\en ,\theta$) is clearly also a  function of zenith angle $\theta$.  In practice,  A($\en ,\theta$) is determined as a function of the incident neutrino direction and zenith angle by a full-detector simulation including the trigger.  It is of the order of 1 km$^2$ for IceCube,  so the effective volume for showers is  V$_{eff}\approx$ A($\en ,\theta)$ L($\theta$)$\approx$ 2 km$^3$.  Finally the expected event rate is 
\be
N_{ev}\approx T \rho_{ice}\,N_A\, V_{eff} \int_{E_{th}}^\infty   \sigma_{\nu N}(\en)\, \frac{dN_\nu}{d\en}\,d\en,
\label{evneu1}
\ee
where E$_{th}$  is the threshold energy,   $ \sigma_{\nu N}(\en)=5.53\times 10^{-36}(\en/GeV)^{0.363}$ cm$^2$ is the charged current cross section \citep{1998PhRvD..58i3009G}.\\
Proposing that the neutrino spectrum can be written as
\be
\frac{dN_{\nu}}{d\en}=A_{\nu} \, \left(\frac{\en}{\mbox{GeV}}\right)^{-\alpha_{\nu}},
\label{espneu1}
\ee
where the normalization factor, A$_{\nu}$,  is calculated by correlating the neutrino flux luminosity with the GeV photon flux  \citep{2008PhR...458..173B}.  This correlation is given by 
\be
\int \frac{dN_{\nu}}{d\en}\,\en\,d\en=\int \frac{dN_\gamma}{dE_\gamma}\,E_\gamma\,dE_\gamma\,.
\ee
Here,  we have used $K=1$ for $p\,p$ interactions \citep[see, e.g.] [and reference therein]{2007Ap&SS.309..407H}. Assuming that the spectral indices for neutrino and $\gamma$-ray spectrum are similar  $\alpha\simeq \alpha_\nu$ \citep{2008PhR...458..173B}, and taking into account that  each neutrino  carries 5\%  of the  proton energies ($\en=1/20\,E_p$)  and  each photon carries  16.7\% of proton energy \citep{Halzen:2013bta}, then the  normalization factors are related by
\be
A_{\nu}=A_{pp,\gamma}\,\left (10\,\xi\right)^{-\alpha+2}\, {\rm GeV}^{-2},
\ee
where A$_{pp,\gamma}$ is given by Eq. (\ref{App}).\\
If we assume that the neutrino spectrum  extends continually over the whole energy range \citep{2008PhRvD..78b3007C}, then  the expected number of neutrinos is
\be
N_{ev} \approx  \frac{T \rho_{ice}\,N_A\, V_{eff}}{\alpha-1.363}\,A_{\nu}\,(5.53\times 10^{-36}\,{\rm cm^2})\left(\frac{E_{\nu,th}}{{\rm GeV}}\right)^{\alpha+1.363}\,{\rm GeV}^{-1}.
\label{numneu1}
\ee
and if it is broken at energy $E_{\nu,b}=\frac{1}{20} E_{p,b}$, then 
\begin{equation}
\label{espneu2}
\frac{dN_\nu}{dE_\nu}   = A_{\nu}
\cases {
(\frac{E_\nu}{\,{\rm GeV}})^{-\alpha} 						& 	$E_\nu < E_{\nu,b}$,\cr
(\frac{E_{\nu,b}}{\,{\rm GeV}})^{-\alpha+\beta} (\frac{E_\nu}{\,{\rm GeV}})^{-\beta}          & 	$E_{\nu,b} \leq  E_\nu$.\cr
}
\end{equation}
For this case,  the expected event is
\bary
N_{ev}&\approx& \frac{T \rho_{ice}\,N_A\, V_{eff}}{\alpha-1.363}\,A_{\nu}\,(5.53\times 10^{-36}\,{\rm cm^2})\cr
&&\times\left[ \left(\frac{E_{\nu,th}}{{\rm GeV}}\right)^{-\alpha+1.363} + \frac{\alpha-\beta}{\beta-1.363}\,\left(\frac{E_{\nu,b}}{{\rm GeV}}\right)^{-\alpha+1.363}   \right]\,{\rm GeV}^{-1} .
\label{numneu2}
\eary
where the higher spectral index, $\beta$, is given by the broken  power law of proton spectrum. 
\section{Analysis and Results}
\begin{figure}
{\centering
\resizebox*{0.57\textwidth}{0.37\textheight}
{\includegraphics{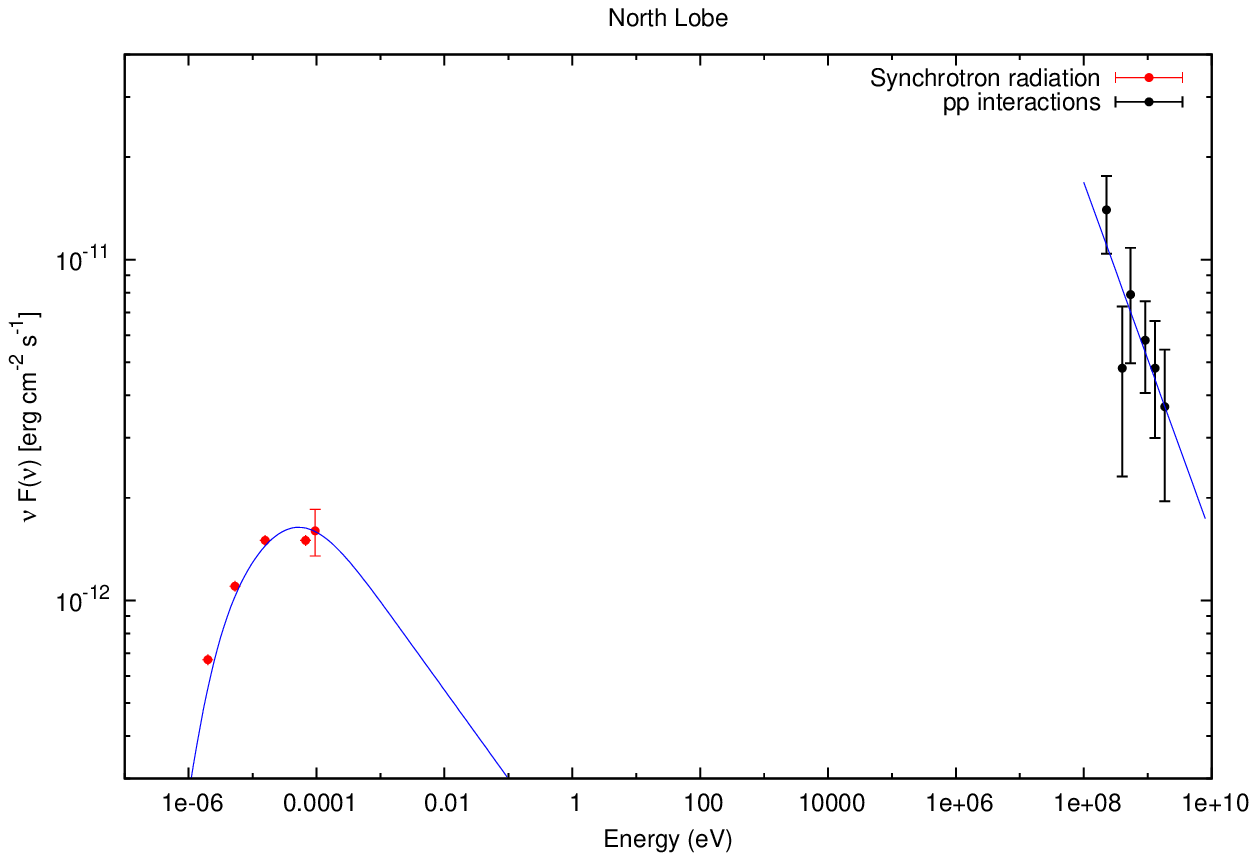}}
\resizebox*{0.57\textwidth}{0.37\textheight}
{\includegraphics{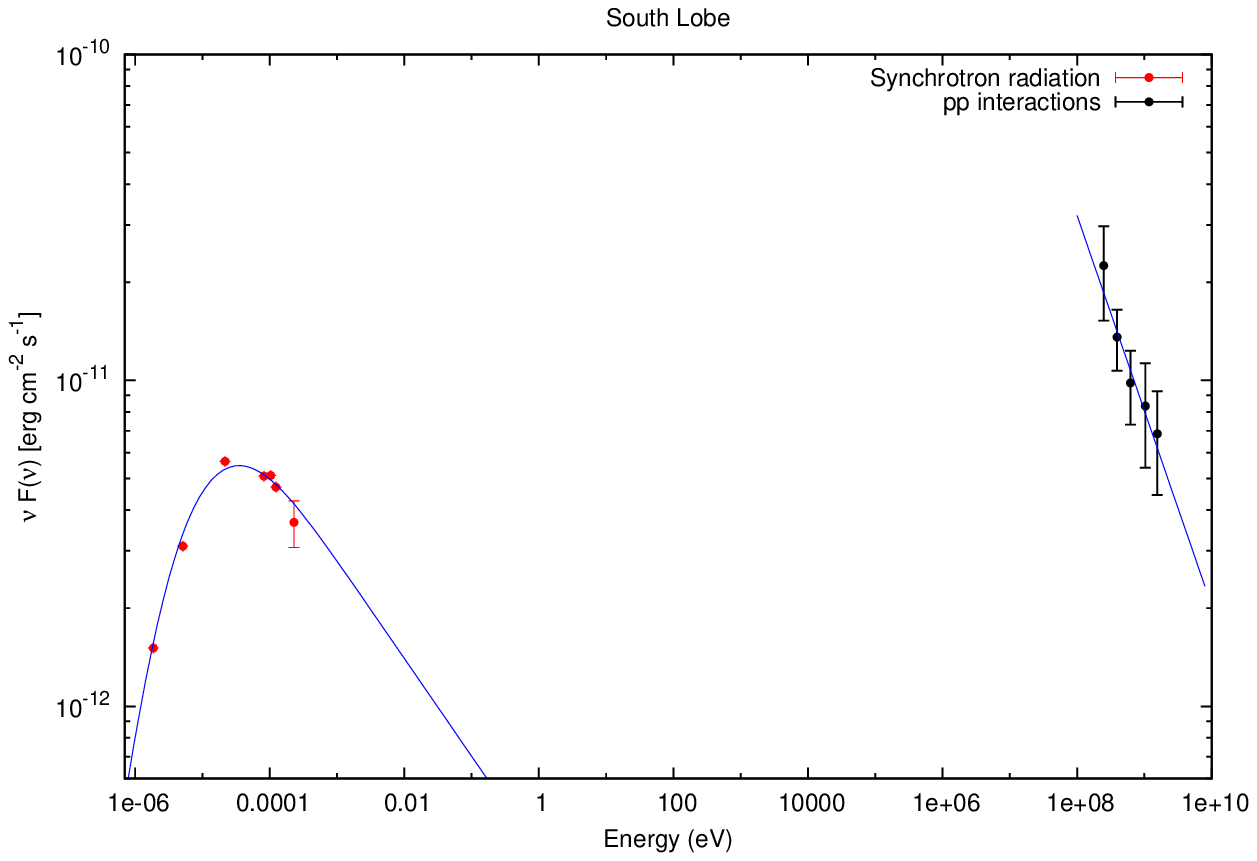}}
}
\caption{Fit of observed SED of the north (left) and south (right) lobes of Cen A.  The peak at  radio wavelength is described  using synchrotron radiation and the $\gamma$-ray spectrum is explained through pp interactions.}
\label{sed}
\end{figure}

We have developed  a synchrotron emission and pp interaction model  to describe the spectra for the north and south lobes of Cen A.  In the synchrotron radiation model we have used an electron distribution described by  a broken power law (eq. \ref{espele}) with the minimum Lorentz factor calculated through electron density and  electron energy density  (eq. \ref{lorentz}). The synchrotron spectrum obtained (eq. \ref{espsyn}) depends on magnetic and electron energy densities,  electron density, size of emission region and cooling time scale characteristic for this process, through the synchrotron normalization constant (eq. \ref{Asyn}) and break energies (characteristic and cut off) (eq. \ref{synrad}).     Taking into account that the jet extends out to  $\sim\,$3 kpc in projection in radio \citep{2007ApJ...670L..81H,2003ApJ...593..169H, 1998AJ....115..960T}, we consider an emission region scale of this size.  Also we have supposed that the magnetic and electron energy densities are equipartitioned through the parameter $\lambda_{e,B}=U_e/U_B$.   In the pp interaction model,  we have considered  accelerated protons described by  a simple power law (eq. \ref{esppr1}) which could be accelerated in the jet or/and the size of the lobe and furthermore interact with  thermal particles in the lobes.  The spectrum generated by this process (eq. \ref{pp}) depends on the proton luminosity (through $A_p$), number density of thermal particles,   age of the lobe  and  spectral index. The age of the lobe can be estimated through cooling time scale of radiating electrons \citep{2009MNRAS.393.1041H} and the number density lies in the range $10^{-10} \,{\rm cm}^{-3} \leq  n_p  \leq 10^{-4}\, {\rm cm}^{-3}$.\\
To find the best fit of the set of model parameters  with data for each lobe, we use the method of Chi-square ($\chi^2$) minimization \citep{1997NIMPA.389...81B}.  We have fitted  WMAP and Fermi  data for each of the lobes with synchrotron emission (eqs. \ref{synrad} and \ref{espsyn})  and pp interaction (eq. \ref{pp}), respectively.  As shown in appendix A,  firstly  we found the photon spectral index ($\alpha$) and the normalization constant  ($A_{pp,\gamma}$)  of  the $\gamma$-ray spectrum generated by pp interaction model.  Secondly, with the fitting spectral index we fit WMAP data with the synchrotron model to find  the  break energies (characteristic $\epsilon^{obs}_{\gamma,m}$ and cut off $\epsilon^{obs}_{\gamma,c}$ ) and the normalization constant characteristic of this process ($A_{syn,\gamma}$). Finally, after fitting the SED of each lobe,  we  plot fig. \ref{sed} presenting  also the best set of  these parameters:   pp interaction parameters in table A1 and synchrotron radiation parameters in table A2 (see  appendix A).\\
\noindent As shown in  table 1, it can be seen that the values of normalization constant  $A_{pp,\gamma}$  and  photon spectral index ($\alpha$) of the  $\gamma$-ray  spectra for each of the lobes are [$5.10\pm 0.96$] erg cm$^{-2}$ s$^{-1}$ (north) and [$8.07\pm 1.58$] erg cm$^{-2}$ s$^{-1}$ (south), and 2.519 (north)  and 2.598 (south), respectively,  which were firstly obtained by \citet{2010Sci...328..725A}.
From the values of the best set of parameters obtained with  the synchrotron model  (table A2 and  eqs. \ref{Asyn} and \ref{synrad}),  we plot  the  synchrotron cooling time scale ($t_{syn}$),  the electron density ($N_e$) and equipartition parameter ($\lambda_{e,B}$) as a function of magnetic field ($B$), as shown in fig. \ref{fit_syn}. For the north lobe, considering the value of equipartition parameter $\lambda_{e,B}=4.3$ \citep{2010Sci...328..725A}, we found the value of magnetic field  B=3.41 $\mu$G, and then  the values of  synchrotron cooling time $t_{syn}$=55.1 Myr and  electron density $N_e=2.1 \times 10^{10}\, cm^{-3}$. For the south lobe,  considering the value of equipartition parameter $\lambda_{e,B}=1.8$ \citep{2010Sci...328..725A}, we found the value of magnetic field  B=6.19 $\mu$G, and then  the values of  synchrotron cooling time $t_{syn}$=27 Myr and  electron density $N_e=3.9 \times 10^{10} cm^{-3}$.\\ 
On the other hand, from the values of the observed quantities and parameters given in tables A1 and A2, firstly we  analyze the contributions of p$\gamma$ interactions and inverse Compton scattering  of CMB and EBL to the $\gamma$-ray spectra and secondly, from pp interactions we correlate the $\gamma$-ray, UHECRs and neutrino fluxes to  estimate the number of  UHE protons and neutrinos expected in PAO  and IceCube, respectively. These estimations are done, assuming that these spectra are extended up to the energy range of each of the experiments.\\
Relativistic electrons may upscatter synchrotron photons up to higher energies given by
\be
E^{ic}_{\gamma,k}\simeq  \gamma^2_{e,m/c} E_{\gamma,k},
\label{ic}
\ee
where the index $k$ represents the CMB and EBL photons \citep{2001ARA&A..39..249H,  2008ApJ...686L...5G, 2008IJMPD..17.1515R, 2013avhe.book..225D}.  Taking into account the radiation power typical of this process $dE/dt=4/3 \sigma_T\beta^2\gamma^2_e\,U_\gamma$ and  performing a process similar to that done with synchrotron emission, the IC spectrum can be written as
\begin{equation}
\label{espinv}
{\epsilon^{ic}_\gamma}^2 N_\gamma(\epsilon^{ic}_\gamma) = A_{ic,\gamma}
\cases {
(\frac{\epsilon^{ic}_\gamma}{\epsilon^{ic}_{\gamma,m}})^{4/3}    &  $\epsilon^{ic, obs}_\gamma < \epsilon^{ic, obs}_{\gamma,m}$,\cr
 (\frac{\epsilon^{ic}_\gamma}{\epsilon^{ic}_{\gamma,m}})^{-(\alpha-3)/2}  &  $\epsilon^{ic,obs}_{\gamma,m} < \epsilon^{ic,obs}_\gamma < \epsilon^{ic,obs}_{\gamma,c}$,\cr
(\frac{\epsilon^{ic}_{\gamma,c}}{\epsilon^{ic}_{\gamma,m}})^{-(\alpha-3)/2}    (\frac{\epsilon^{ic}_\gamma}{\epsilon^{ic}_{\gamma,c}})^{-(\alpha-2)/2},           &  $\epsilon^{ic,obs}_{\gamma,c} < \epsilon^{ic,obs}_\gamma < \epsilon^{ic, obs}_{\gamma,max} $\cr
}
\end{equation}
where
\be
\label{Aic}
 A_{ic,\gamma}=  \simeq \frac{\sqrt{8\pi} \sigma_T\,\beta^2}{9\,q_e\,D_z^2}(1+z)^{-1}\,\delta_D\,U_B^{-1/2}\,N_e\,r_d^3\,U_\gamma\,E_{\gamma,k},
\ee
and $U_\gamma$ is the photon energy density of  CMB and EBL.    Replacing the electron Lorentz factors and the typical photon energies of CBM and EBL \citep{2013avhe.book..225D} (eqs. (\ref{synrad}) and (\ref{lorentz})) in eqs. (\ref{ic}),  (\ref{espinv}) and (\ref{Aic}) and from the best set of parameters,  we plot these contributions, as shown in fig. \ref{fit_ic}.  In this figure, we can notice that a superposition of inverse Compton of CMB and EBL, and pp interactions could describe satisfactorily the observed Fermi-LAT fluxes.
 Otherwise, p$\gamma$  interactions take place when accelerated protons collide with  target photons.  The single-pion production channels are $p+\gamma\to n+\pi^+$ and $p+\gamma\to p+ \pi^0$, where the relevant pion decay chains are $\pi^0\to 2\gamma$, $\pi^+\to \mu^++\nu_\mu\to e^++\nu_e+\bar{\nu}_\mu+\nu_\mu$ and $\pi^-\to \mu^-+\bar{\nu}_\mu\to e^-+\bar{\nu}_e+\nu_\mu+\bar{\nu}_\mu$. Taking into account  that $\pi^0$  carries $20\%$ of the proton's energy and that each produced photon shares the same energy then, the observed HE photon is given by
\be
E^{obs}_{\gamma,HE}\simeq 0.5\frac{\delta^2_D\,(m^2_\Delta-m_p^2)}{(1+z)^2}(E^{obs}_{\gamma,LE})^{-1},
\label{pgamma}
\ee
where $E^{obs}_{\gamma,LE}$ corresponds to low-energy (LE) photons. Based on eq. (\ref{pgamma}), it is necessary that target photons should be in the energy range of $E_{\gamma,LE}\sim$ (30 - 460) MeV for a full description of  $\gamma$-ray spectra. Although a more robust analysis  should be done such as  a calculation of density of target photons, optical depth,  rate of energy loss, etc, a simple calculation shows that this process needs seed photons with energies  from tens to hundreds of MeV which is completely different to  $\gamma$-ray spectra observed by LAT.  Hence, there is no contribution of p$\gamma$ emission  to the $\gamma$-ray spectra.\\
On the other hand, in addition to the analysis performed and showed here about the SED of the lobes, we are going to see whether there is any correlation of $\gamma$-ray spectra and  UHECRs collected with PAO.   Estimating the age of the lobes by means of synchrotron cooling time, $t_{lobe}\simeq t_{syn}$ \citep{2009MNRAS.393.1041H} and the acceleration (eq. \ref{tac})   and  diffuse  (eq. \ref{tac})  time  scales with the magnetic field found we obtain that the acceleration time,  diffuse time and the age of the lobes are  0.65 (0.37) Myr,  0.49 (0.89) Myr and 55.1 (27) Myr for the north (south)  lobe, respectively.   Comparing the time scales,   $t_{acc}\sim t_{diff}\ll t_{lobe}$, one can calculate  that the maximum proton energies required  are $E_{p,max}= 8.67 (15.81)\times 10^{19} $ eV for the north (south) lobes, hence  we demonstrate that protons could be accelerated up to  energies as high as 10$^{20}$ eV in the lobes.  Following our analysis, we  replace  the fitting parameters t$_{lobe}$, $\alpha$ and $A_{pp,\gamma}$ in eqs. \ref{nUHE1} and  \ref{lum} to calculate the proton luminosity from  $\sim$ GeV to 10$^{11}$ GeV and also estimate the number of UHE protons expected with PAO.  For this calculation we take into account  the thermal number density in the range  $10^{-10} \,{\rm cm}^{-3} \leq  n_p  \leq 10^{-4}\, {\rm cm}^{-3}$.  As shown in fig. \ref{prot_lum}, one can see that  proton luminosity increases as thermal density decreases and it decreases as energy increases.   In this plot there are two interesting ranges, the GeV and EeV  ranges. In GeV range which is connected directly with the Fermi fluxes,  the proton luminosities in the north(south) lobe at some GeV energy  are $ 3.7 (7.7) \times 10^{43}$ erg s$^{-1}$ and $ 3.7 (7.7) \times 10^{49}$ erg s$^{-1}$ for minimum ($n_p=10^{-4} {\rm cm}^{-3}$) and maximum  ($n_p=10^{-10} {\rm cm}^{-3}$) thermal particle densities considered and  in the EeV range the luminosities are  $ 3.11 (2.31) \times 10^{38}$ erg s$^{-1}$  and $ 3.11 (2.31) \times 10^{44}$ erg s$^{-1}$ for the same thermal particle densities.
The values of luminosity at GeV energy range are of the same order as  those found by  \citet{2009MNRAS.393.1041H, 2013A&A...558A..19W, 2010Sci...328..725A} and at EeV energy range it is also in accordance with \citet{2009MNRAS.393.1041H, 2009NJPh...11f5016D}.  In fig. \ref{N_uhecr}, the number of events as a function of thermal density is plotted when  a power law was considered.  As shown,  the expected number  decreases as thermal particle density decreases,  reaching a minimum value of 3.52 (north lobe) and 3.0 (south lobe) with density  is equal to $10^{-4} {\rm cm}^3$, see table 1.
\begin{center}\renewcommand{\arraystretch}{0.7}\addtolength{\tabcolsep}{-1pt}
\begin{tabular}{c c c c c c c c}
  \hline \hline
 &  & \scriptsize{North Lobe} &  \scriptsize{South Lobe}  &\\
 \hline 
& \scriptsize{$ n_p (cm^{-3})$} & \scriptsize{Number of UHECRs} &      \scriptsize{Number of UHECRs} & \\
\hline
& \scriptsize{$10^{-4}$}     & \scriptsize{$3.52 $}                              &  \scriptsize{$3.00$} &\\
& \scriptsize{$10^{-5}$}     & \scriptsize{$35.2$}                               &  \scriptsize{$30\times 10^{-2}$} &\\
& \scriptsize{$10^{-6}$}     & \scriptsize{$352 $}                               &  \scriptsize{$300$} &\\
& \scriptsize{$10^{-7}$}     & \scriptsize{$3.52\times 10^{3}$}          &  \scriptsize{$3.00\times 10^{3}$} &\\
& \scriptsize{$10^{-8}$}     & \scriptsize{$3.52\times 10^{4}$}          &  \scriptsize{$3.00\times 10^{4}$} &\\
& \scriptsize{$10^{-9}$}     & \scriptsize{$3.52\times 10^{5}$}          &  \scriptsize{$3.00\times 10^{5}$} &\\
& \scriptsize{$10^{-10}$}   & \scriptsize{$3.52\times 10^{6}$}          &  \scriptsize{$3.00\times 10^{6}$} &\\
\hline
\end{tabular}
\end{center}
\begin{center}
\scriptsize{\textbf{Table 1. Number of expected events (UHECRs)  in PAO as a function of number density of thermal particles  from the north and south lobe of Cen A.}}\\
\end{center}
Considering a simple power law and the number density of thermal particles equal to $10^{-4}\, {\rm cm}^3$, the result of number of expected UHECRs is very similar to that reported by PAO. Also taking into account this number density and from eqs. \ref{pressure} and \ref{Etotal} we  estimate the pressures and energies in the lobes, see table 2.
\begin{center}\renewcommand{\arraystretch}{0.7}\addtolength{\tabcolsep}{-1pt}
\begin{tabular}{c c c c c c c c}
  \hline \hline
 &  & \scriptsize{North Lobe} &  \scriptsize{South Lobe}  &\\
 \hline 
& \scriptsize{ Symbol} & \scriptsize{Values} &      \scriptsize{Values} & \\
\hline
 \scriptsize{Non-thermal pressure (dyn cm$^{-2}$ )}  &  \scriptsize{ P$_{nth}$ } & \scriptsize{$3.5 \times 10^{-12}$}                              &  \scriptsize{$5.4 \times 10^{-12}$} &\\
 \scriptsize{Thermal pressure (dyn cm$^{-2}$ ) }  &  \scriptsize{$P_{th}$}  & \scriptsize{$0.9 \times 10^{-13}$}                               &  \scriptsize{$0.9 \times 10^{-13}$} &\\
 \scriptsize{Proton density energy (erg cm$^{-3}$ )}  &  \scriptsize{$U_p $}  & \scriptsize{$1.06\times 10^{-12}$}          &  \scriptsize{$1.07\times 10^{-12}$} &\\
 \scriptsize{Total energy (erg)}   &  \scriptsize{$E_{tot}$} & \scriptsize{$4.4\times 10^{59}$}                               &  \scriptsize{$6.6\times 10^{59}$} &\\
\hline
\end{tabular}
\end{center}
\begin{center}
\scriptsize{\textbf{Table 2.  Distributions of pressures and energies in  the north and south lobe of Cen A.  These values have been obtained for  a volume $1.3\times 10^{71}$ cm$^{-3}$}, number density of thermal particles  $n_p\sim 10^{-4}\, {\rm cm}^3$ and temperature $\sim 10^7$ K.}\\
\end{center}
As shown in table 2, the bigger contribution of pressure  exerted on the lobes comes from non-thermal particles and  the contribution of protons to pressure and energy is although not dominant if significant.\\    
For this density and eq. (\ref{numneu1}), we calculate the number of neutrinos expected per year in IceCube  for neutrino threshold energy equal to $E_{\nu,th}$= 1 TeV (see table 3).  In this table, we can see that non neutrinos are expected from the lobes. 
 \begin{center}\renewcommand{\arraystretch}{0.7}\addtolength{\tabcolsep}{-1pt}
\begin{tabular}{c c c c c c c c}
  \hline \hline
 &  & \scriptsize{North Lobe} &  \scriptsize{South Lobe}  &\\
 \hline 
& \scriptsize{$E_{\nu,th}$} & \scriptsize{$N_{ev}/T (year)^{-1}$} &      \scriptsize{$N_{ev}/T (year)^{-1}$} & \\
\hline
& \scriptsize{1 TeV}& \scriptsize{$9.41\times 10^{-2}$}          &  \scriptsize{$7.47\times 10^{-2}$} &\\
 \hline
\end{tabular}
\end{center}
\begin{center}
\scriptsize{\textbf{Table 3. Number of neutrinos expected on IceCube from north and south lobes of Cen A. This number is calculated taking into account the number density of thermal particles and  neutrino threshold energies equal to $n_p=10^{-4} cm^{-3}$ and, $E_{\nu,th}$= 1 TeV, respectively.}}\\
\end{center}
From table 1, one can see that  for any number density below  $10^{-4}$ cm$^{-3}$, the expected UHECRs would be much higher than those reported by PAO, hence these densities should be  excluded  when  a simple power law is considered, but not when we give  careful consideration to a broken proton spectrum.  In other words, taking into account  a number density of less than  $10^{-4}$ cm$^{-3}$, we could expect  more or less events only when a broken proton spectrum is considered (eqs. \ref{esppr2} and \ref{nUHE1} ).  In fig \ref{c_plot},   contour plots of  the  broken spectrum parameters  are plotted,  higher spectral index ($\beta$)  and  break energy ($E_{p,b}$)  as a function of number density for which PAO would expect one  and two events from each of the lobes.    In these graphs can be seen that $\beta$ is higher in the south lobe,  as one event is expected and the number density is  smaller. For instance,  for the break proton energy ($E_{p,b}=2.04\times10^{17}\,eV$) we expect 2 events when $\beta=3.28 (3.29)$ and 1 event when $\beta=3.47 (3.48)$ for $n_p=10^{-5}\,cm^{-3}$  and 2 events when $\beta=3.83 (3.91)$ and 1 event when $\beta=3.99 (4.11)$ for $n_p=10^{-6}\,cm^{-3}$ from north (south) lobe.  From the analysis performed for the proton spectrum  described by  broken power law one can see that it is more favorable when densities of thermal particles are higher.\\ 
Additionally,  for this case we calculate the number of expected neutrinos  in IceCube.  From eq. (\ref{numneu2}), we plot the number of neutrinos as a function of  time for a broken power law as shown in figs. \ref{n_neu1} and \ref{n_neu2}.   Taking into account the parameters ($\beta$ and $E_{\nu,b}$) for one and two events from the north and south lobe, the events per year are reported in tables B1 and B2.  As shown in fig.  \ref{n_neu1} and \ref{n_neu2},  as $\beta$ and $E_{\nu,b}$ increase broken power laws become closer up to be overlapped.  Assuming a threshold energy of $E_{th,\nu}=1 TeV$, it can be seen that less than 0.1 neutrinos are expected in IceCube. 
\section{Summary and conclusions}
\begin{figure}
{\centering
\resizebox*{0.5\textwidth}{0.31\textheight}
{\includegraphics{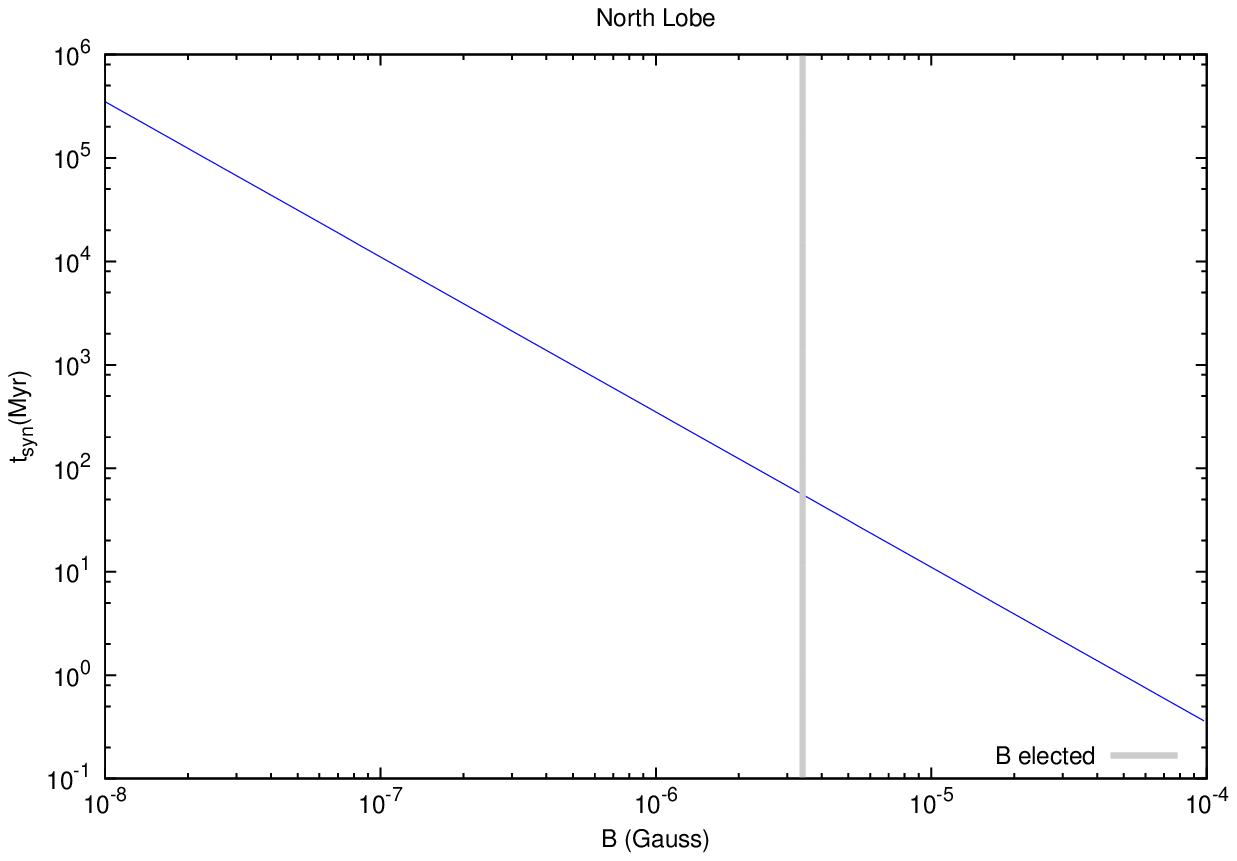}}
\resizebox*{0.5\textwidth}{0.31\textheight}
{\includegraphics{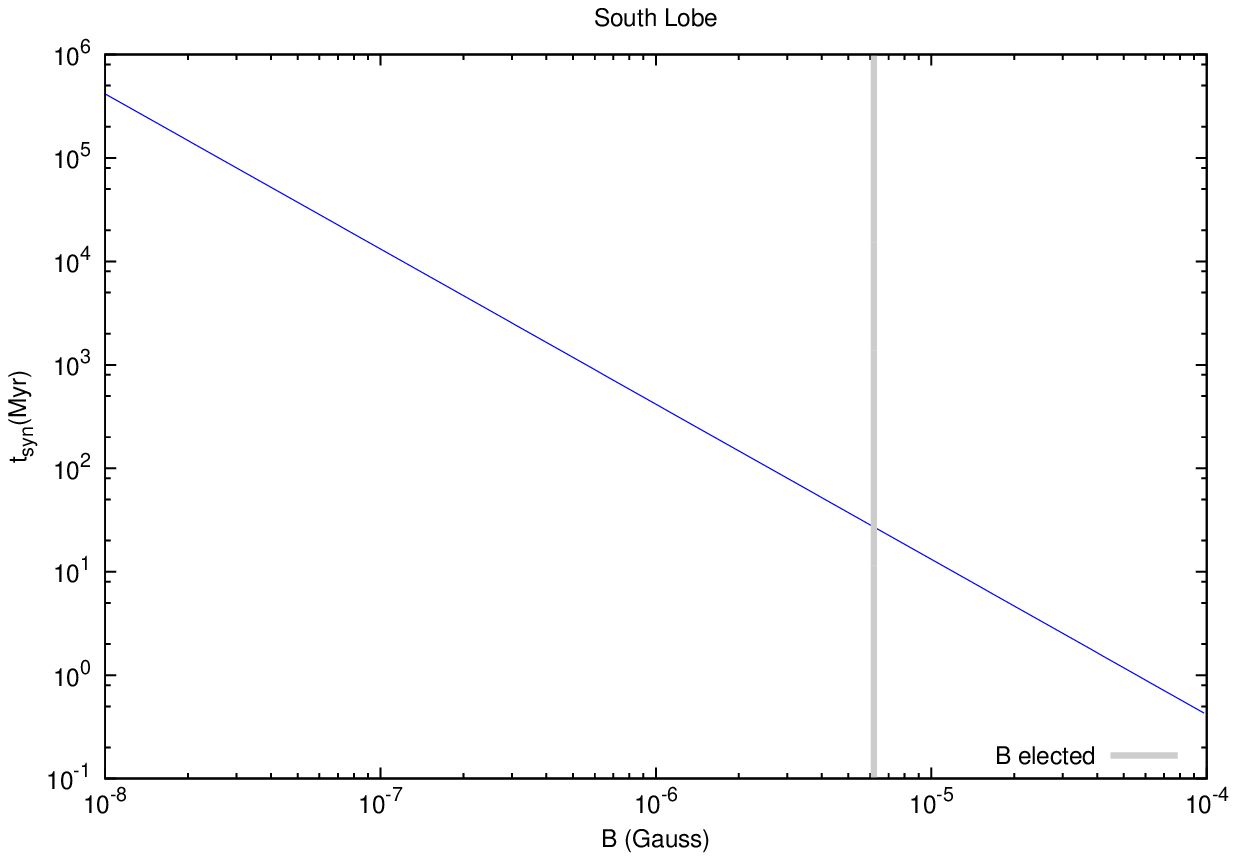}}
\resizebox*{0.5\textwidth}{0.31\textheight}
{\includegraphics{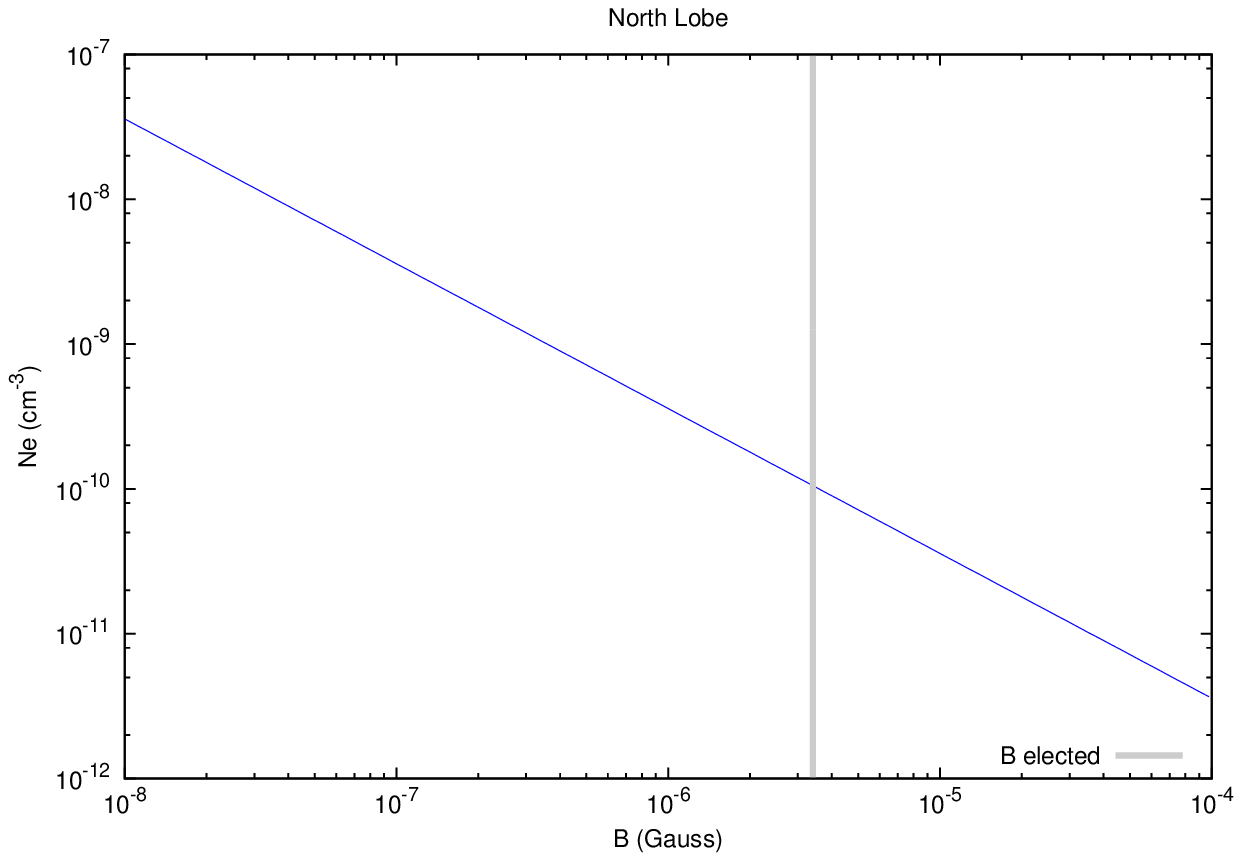}}
\resizebox*{0.5\textwidth}{0.31\textheight}
{\includegraphics{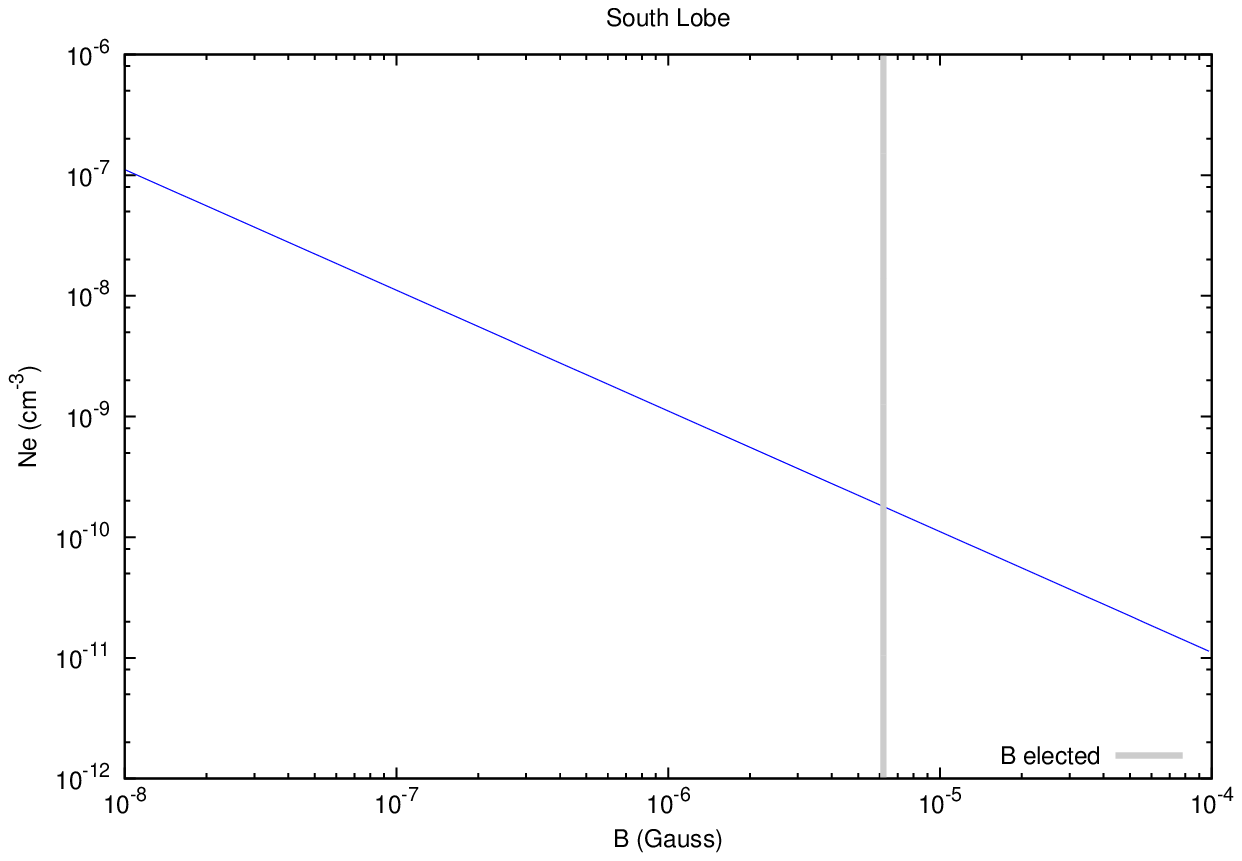}}
\resizebox*{0.5\textwidth}{0.31\textheight}
{\includegraphics{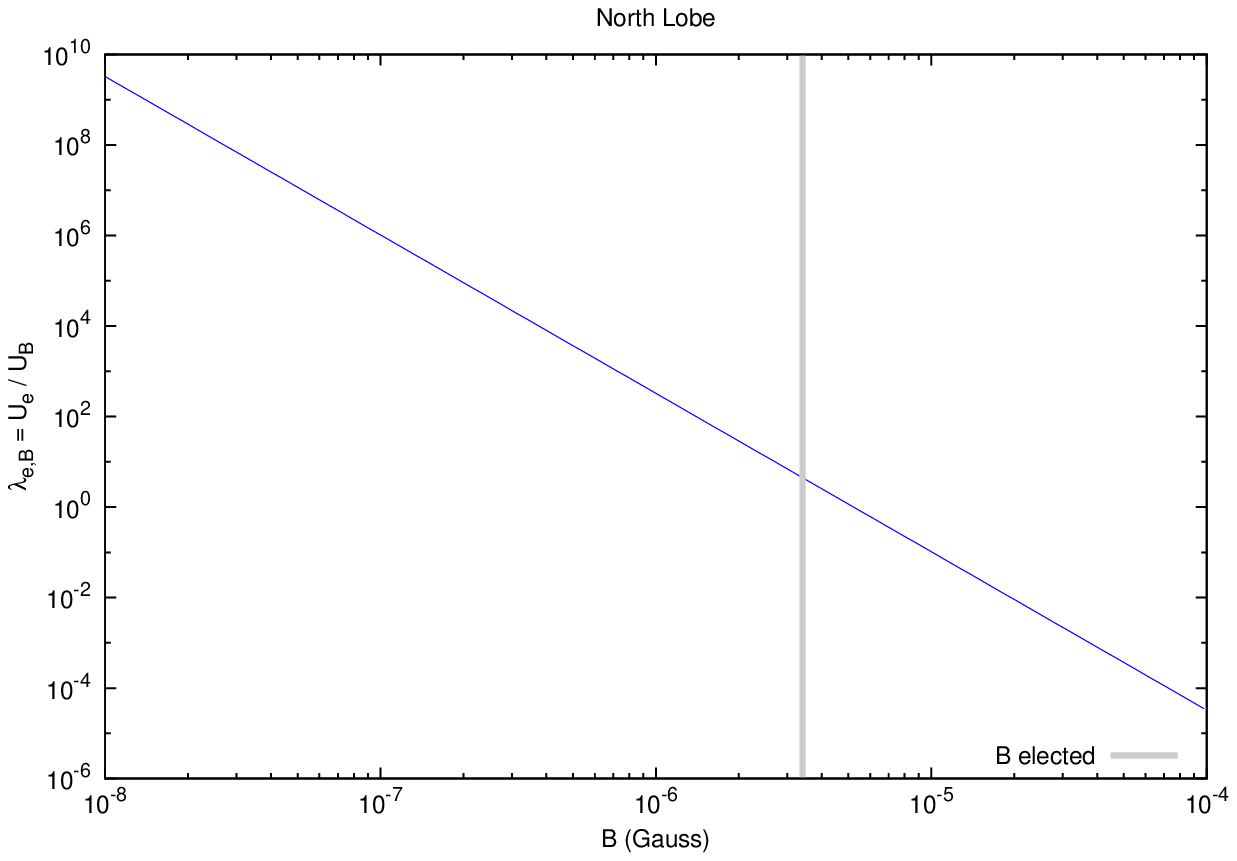}}
\resizebox*{0.5\textwidth}{0.31\textheight}
{\includegraphics{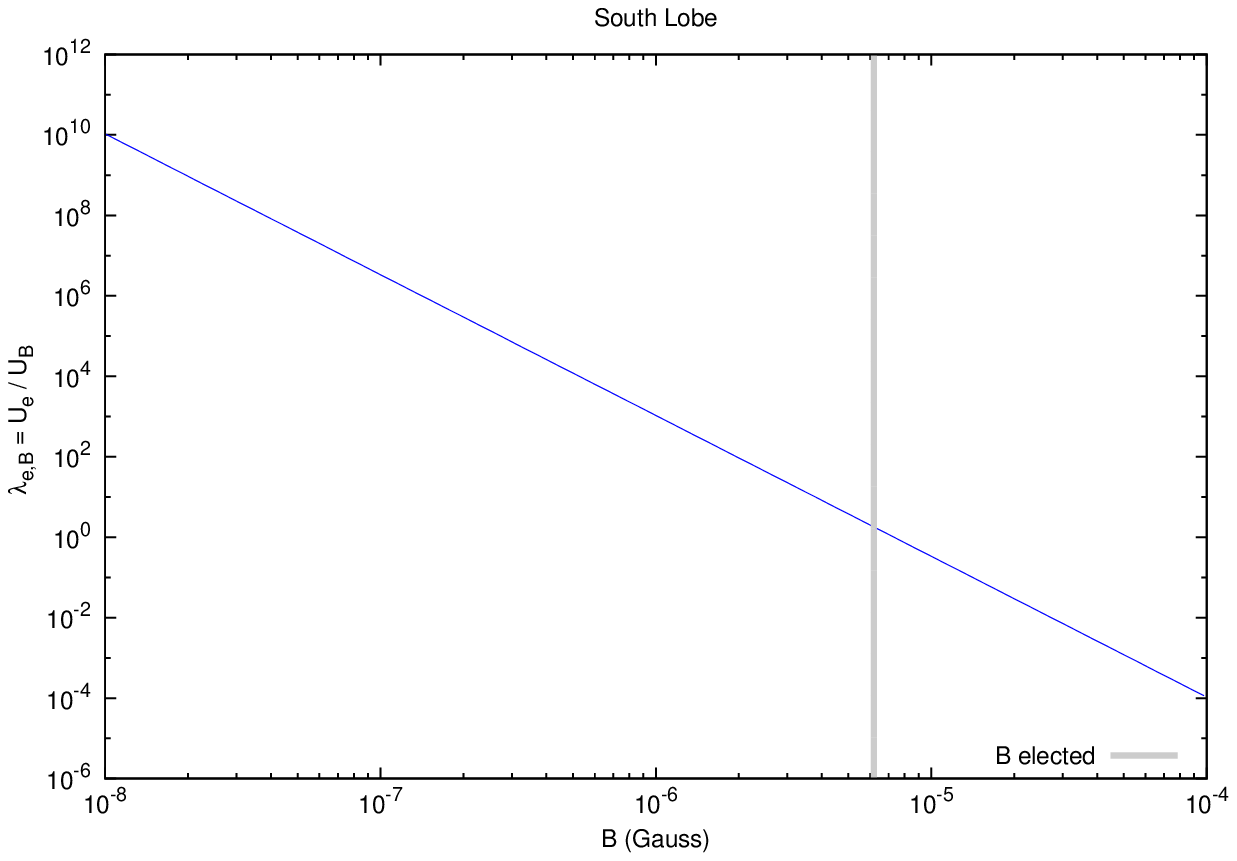}}
}
\caption{Plots of the best set of parameters (synchrotron cooling time ($t_{syn}$), electron density ($N_e$) and equipartition parameter ($\lambda_{e,B}=U_e/U_B$) as a function of magnetic field (B)) obtained with our  synchrotron model for north (left) and south (right) lobes.}
\label{fit_syn}
\end{figure}

\begin{figure}
{\centering
\resizebox*{0.58\textwidth}{0.37\textheight}
{\includegraphics{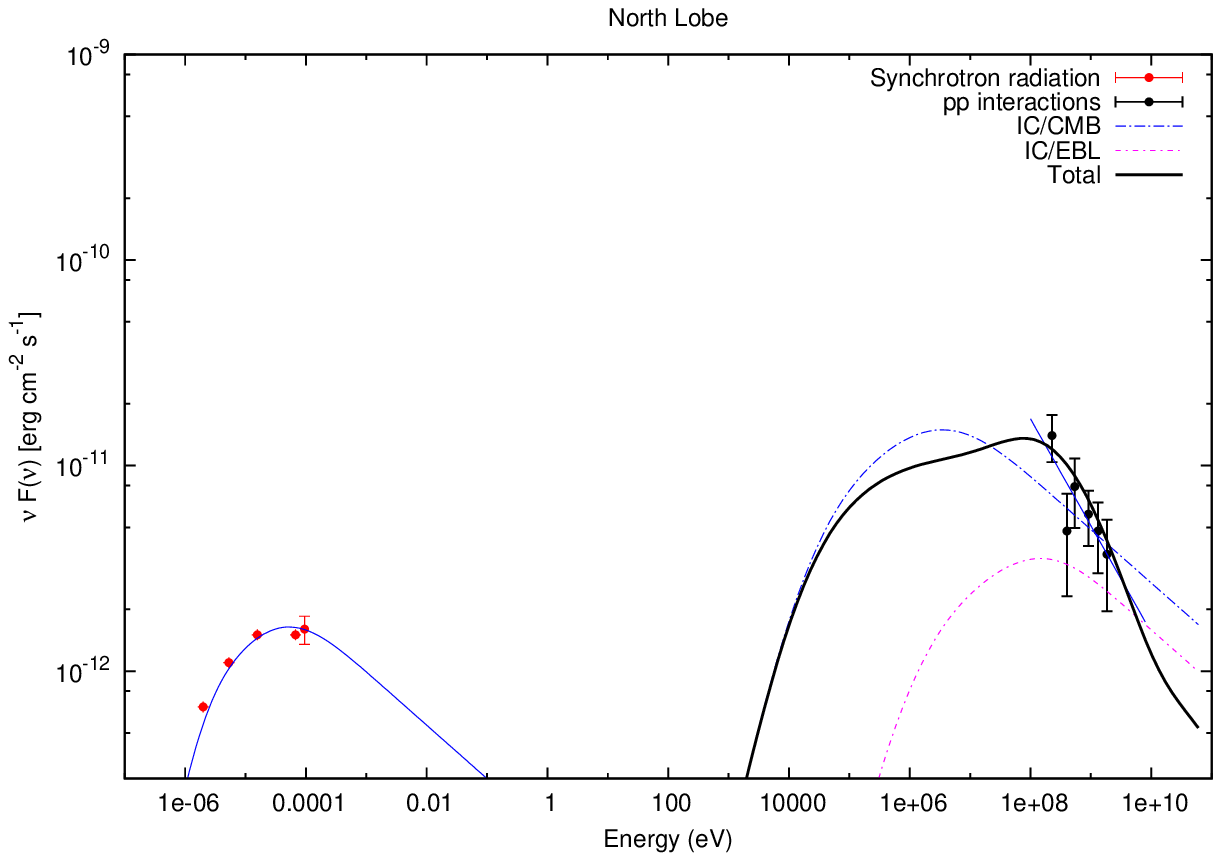}}
\resizebox*{0.58\textwidth}{0.37\textheight}
{\includegraphics{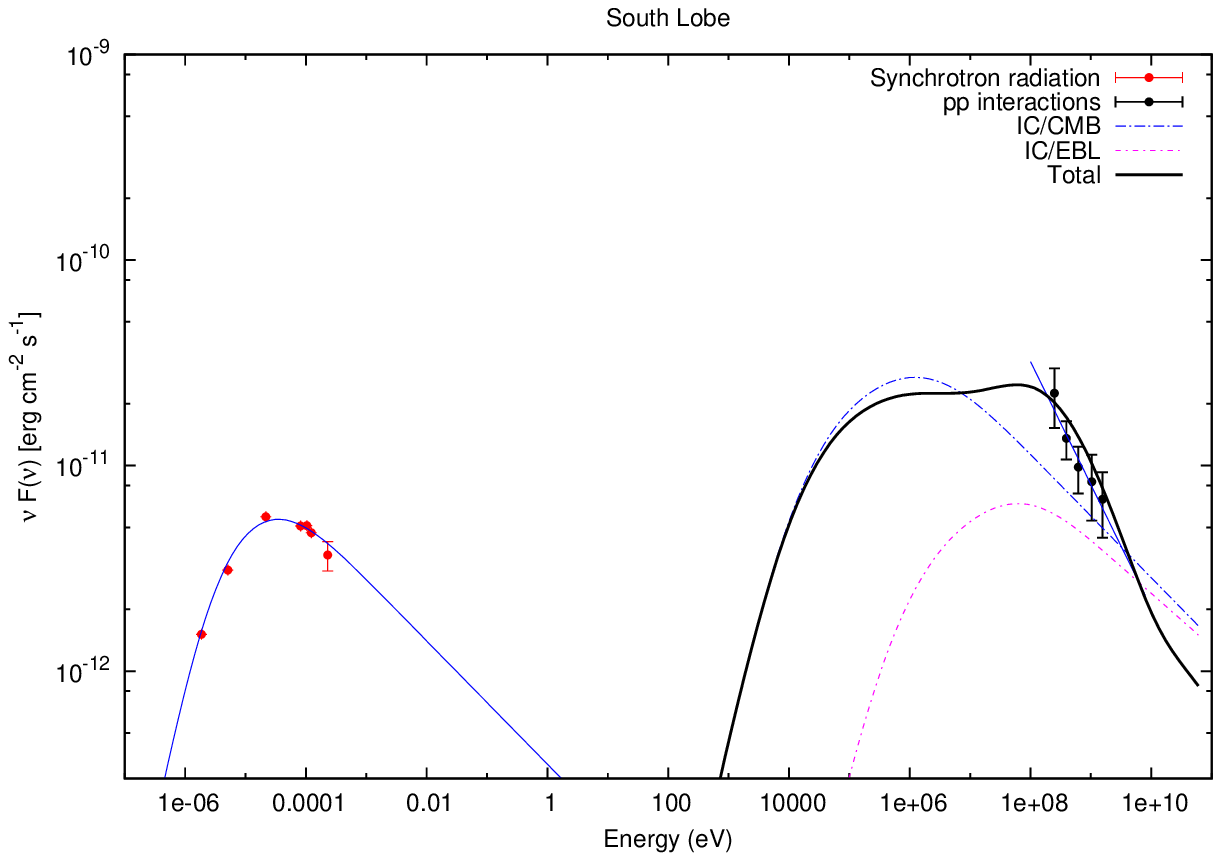}}
}
\caption{Fit of observed SED of the north (left) and south (right) lobes of Cen A.  The component at higher energies is described through pp interaction and  IC (CBM and EBL).  IC emission is  plotted with our model and  the parameters obtained in table 1 and 2.}
\label{fit_ic}
\end{figure}

\begin{figure}
{\centering
\resizebox*{0.58\textwidth}{0.37\textheight}
{\includegraphics{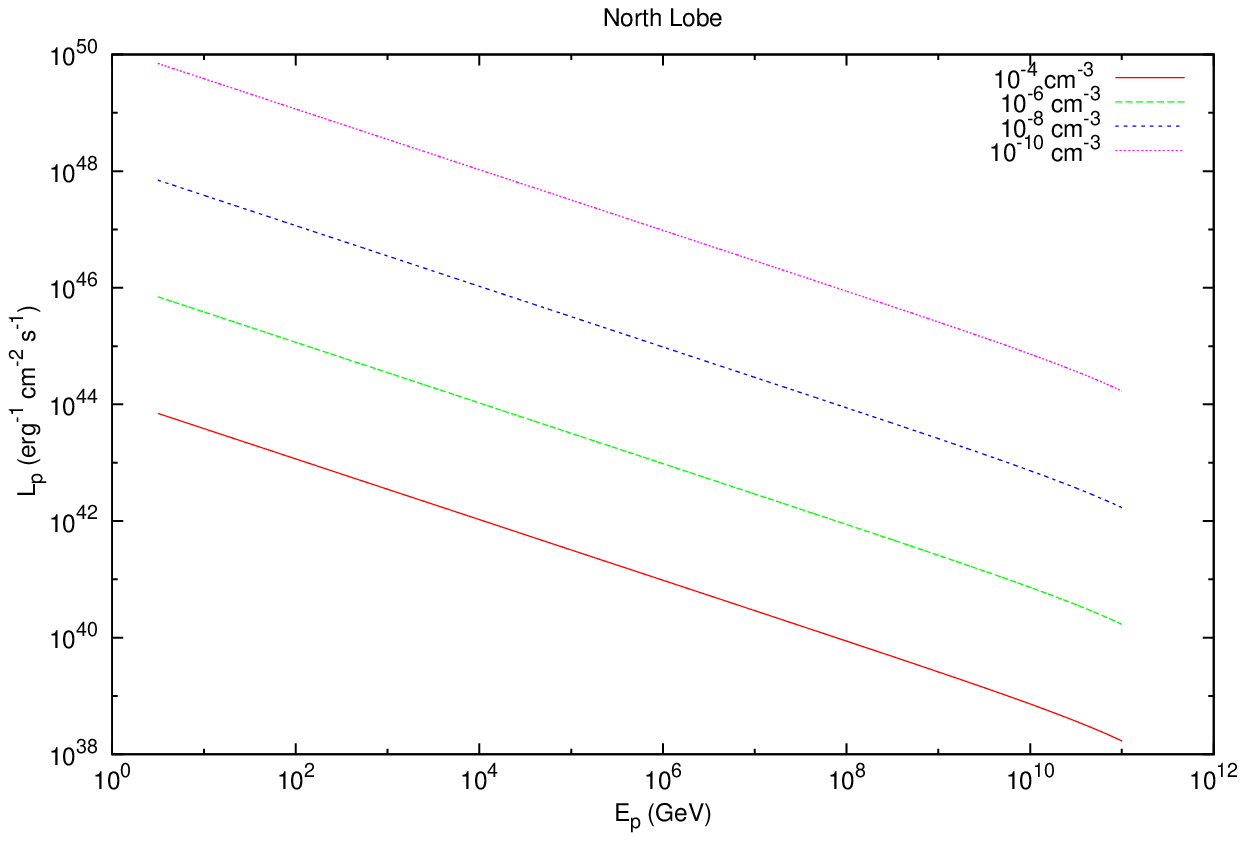}}
\resizebox*{0.58\textwidth}{0.37\textheight}
{\includegraphics{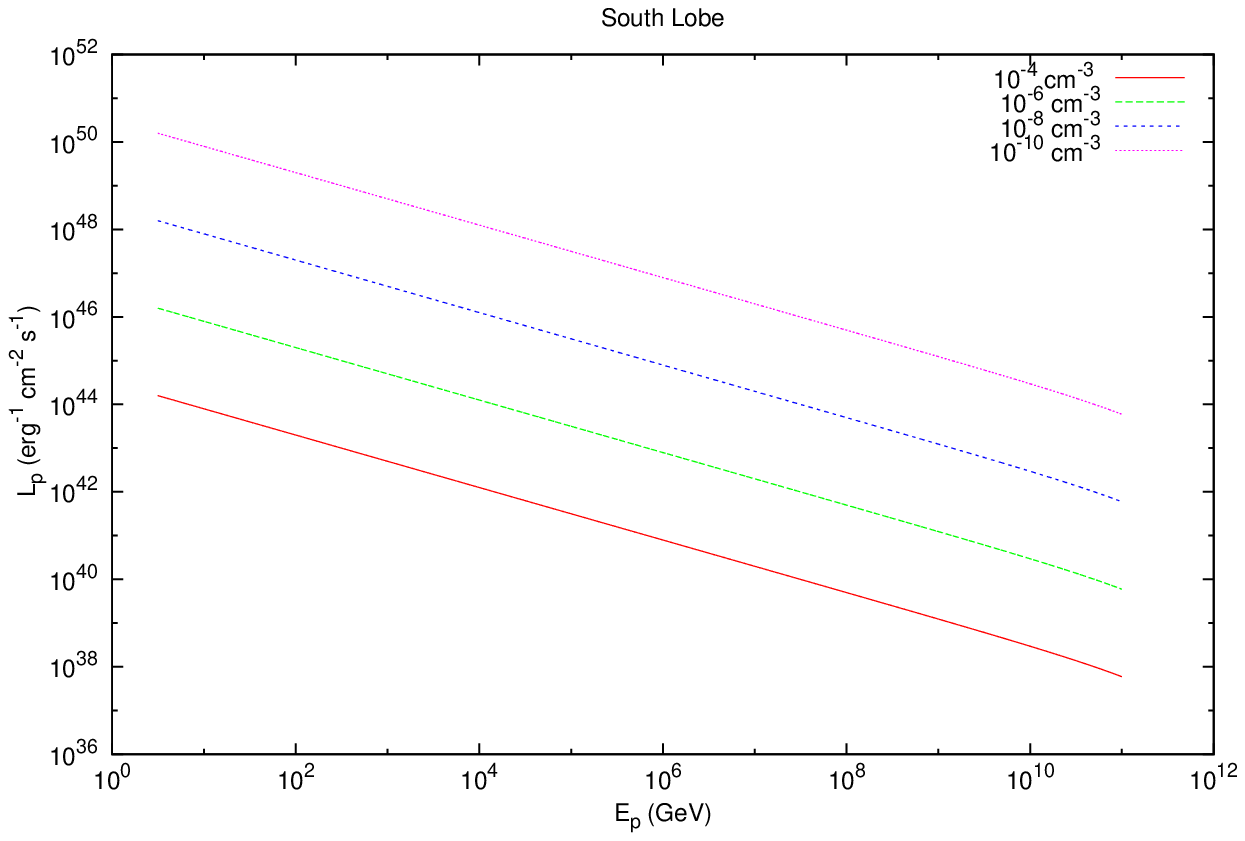}}
}
\caption{Proton luminosity $L_p$ as a function of  energy $E_p$ for   north (left) and south (right) lobes. These plots are generated as
the thermal density range is $10^{-10} \,{\rm cm}^{-3} \leq  n_p  \leq 10^{-4}\, {\rm cm}^{-3}$. }
\label{prot_lum}
\end{figure}

\begin{figure}
\vspace{0.5cm}
{\centering
\resizebox*{0.58\textwidth}{0.37\textheight}
{\includegraphics{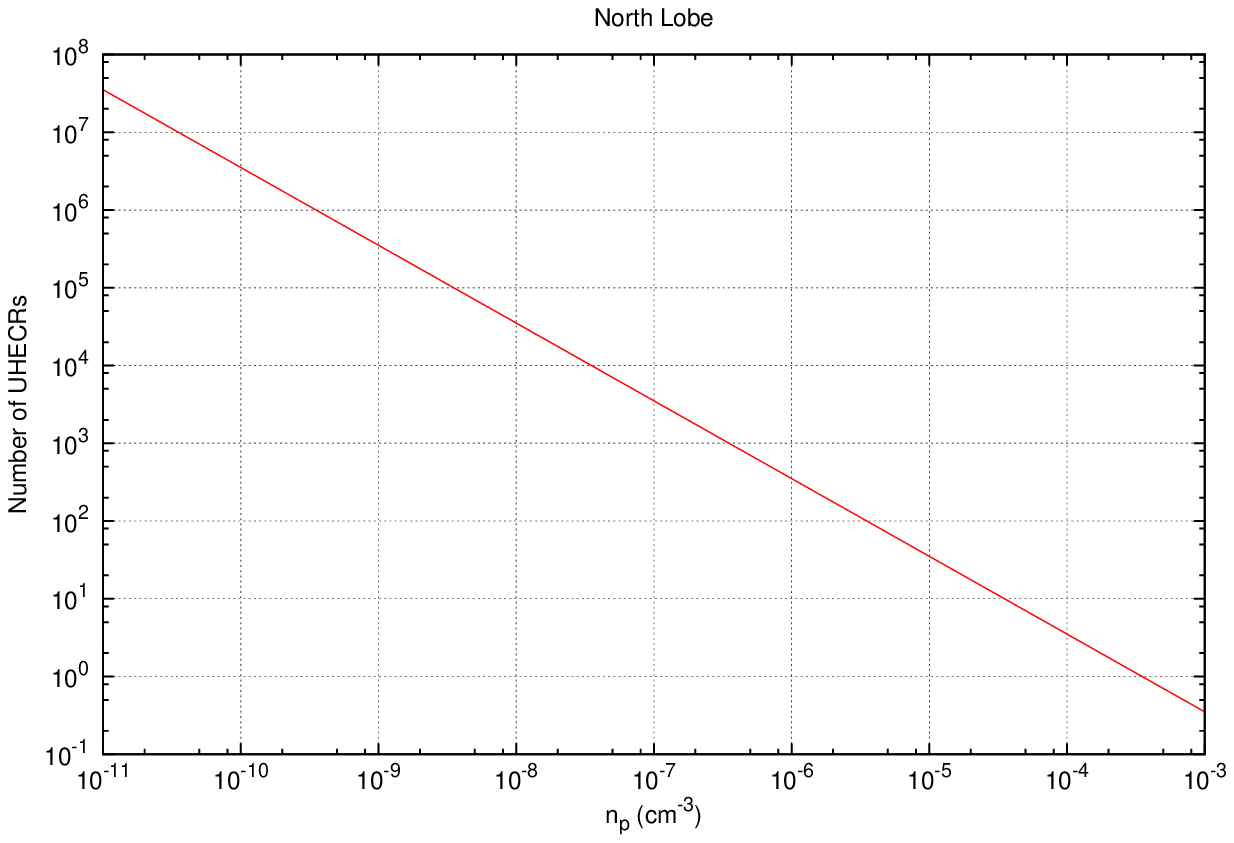}}
\resizebox*{0.58\textwidth}{0.37\textheight}
{\includegraphics{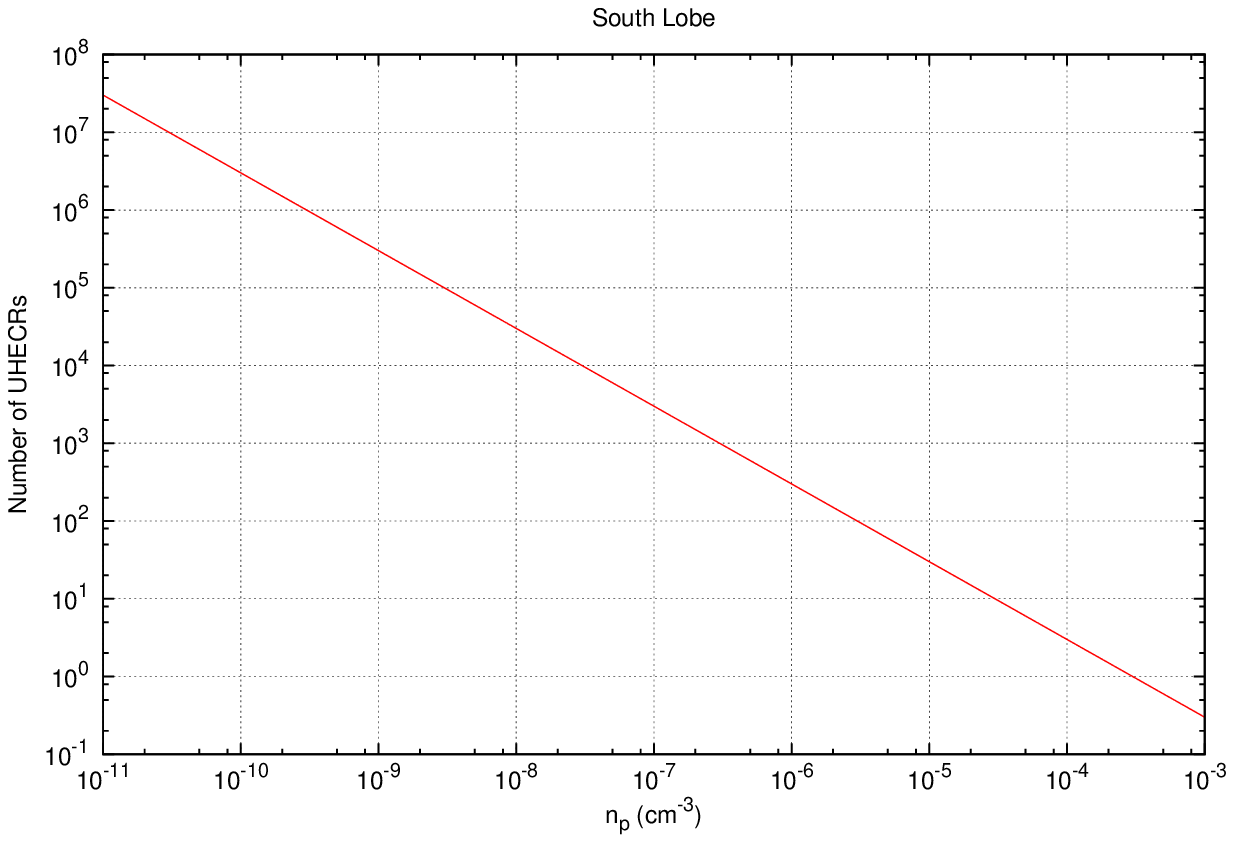}}
}
\caption{Number of UHECRs expected as a function of number density of thermal particles ($n_p$) from the north (left) and south (right) lobes of Cen A.}
\label{N_uhecr}
\end{figure}

\begin{figure} 
{\centering
\resizebox*{0.54\textwidth}{0.38\textheight}
{\includegraphics{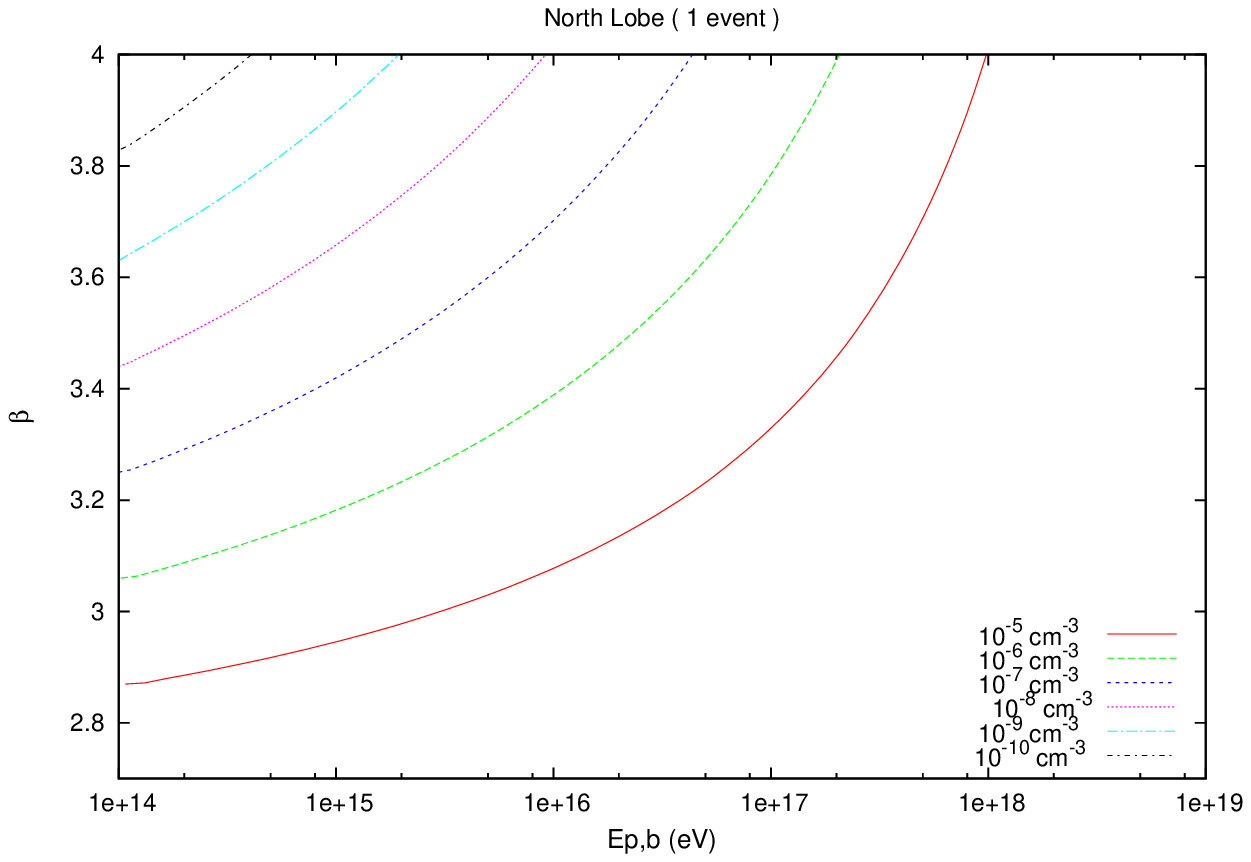}}
\resizebox*{0.54\textwidth}{0.38\textheight}
{\includegraphics{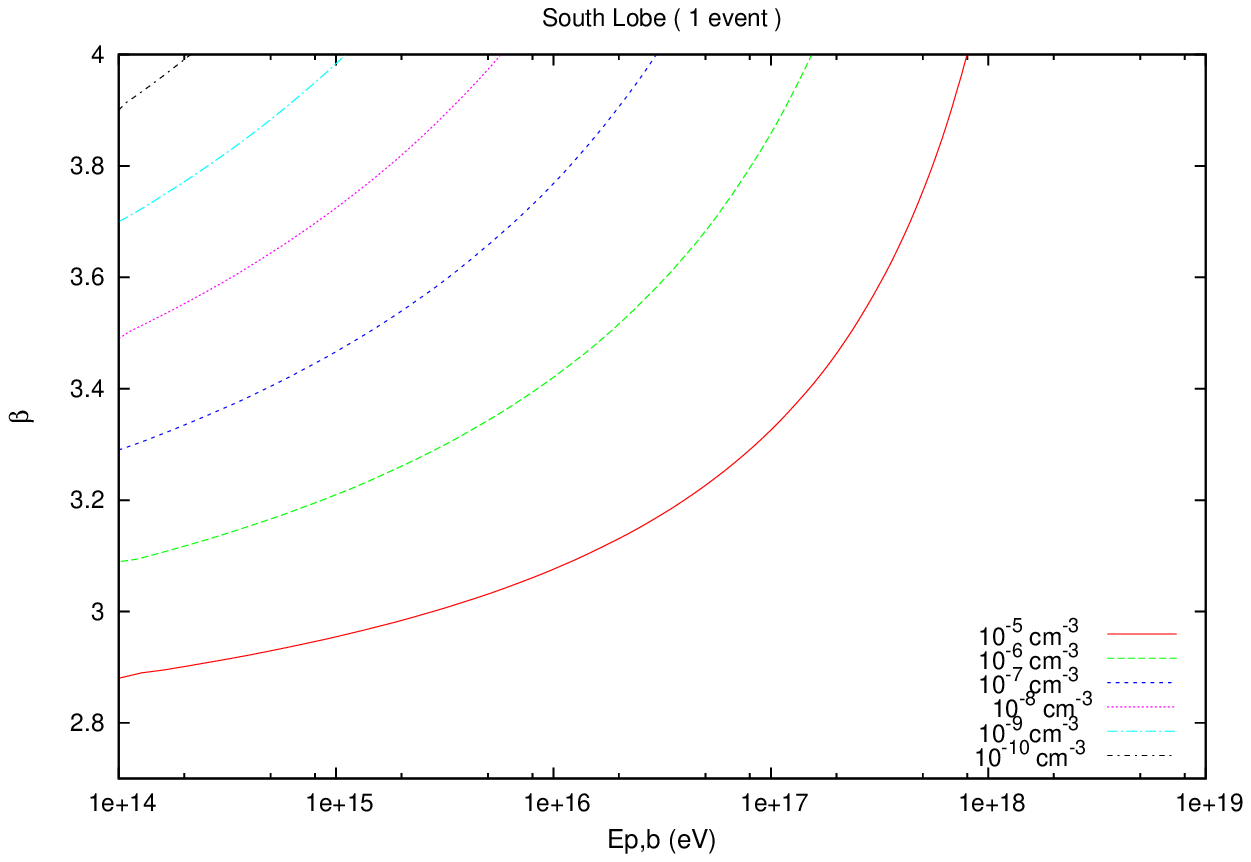}}
\resizebox*{0.54\textwidth}{0.38\textheight}
{\includegraphics{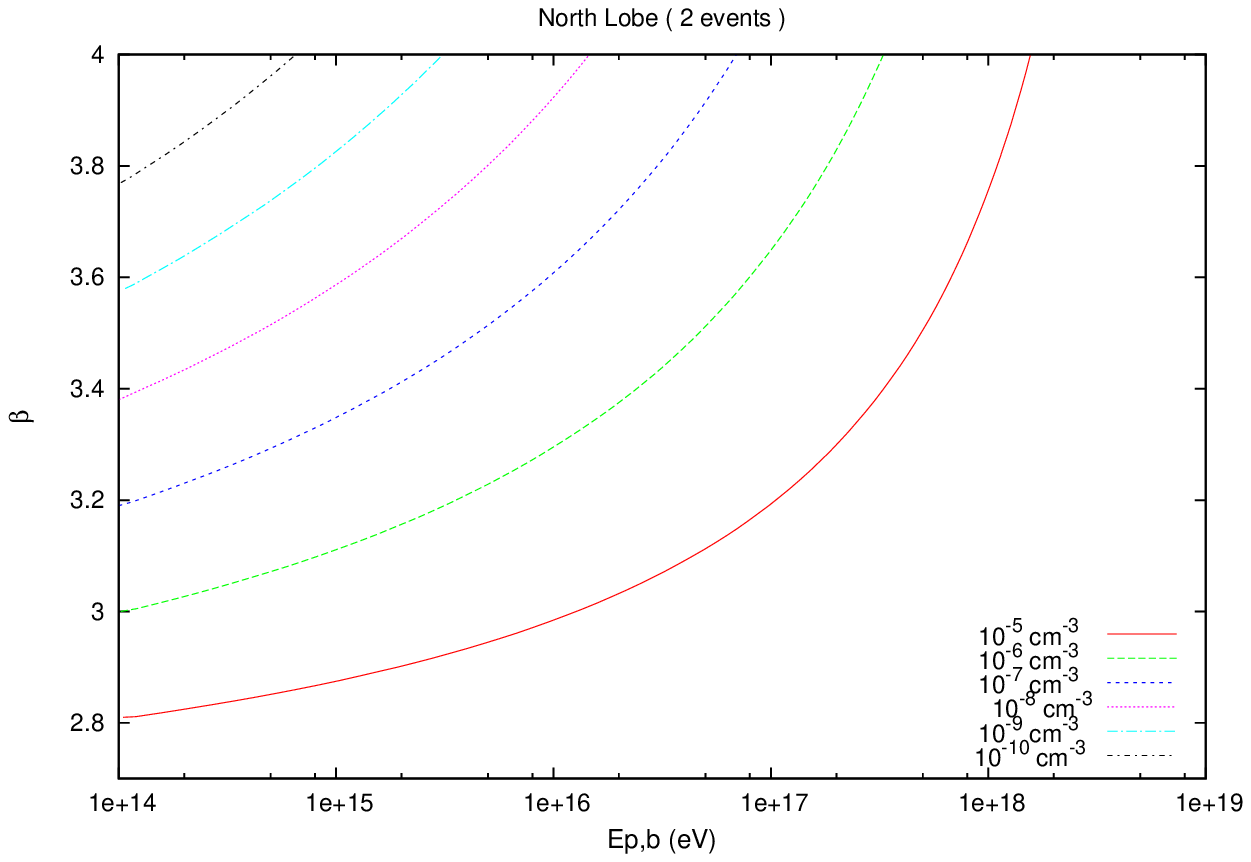}}
\resizebox*{0.54\textwidth}{0.38\textheight}
{\includegraphics{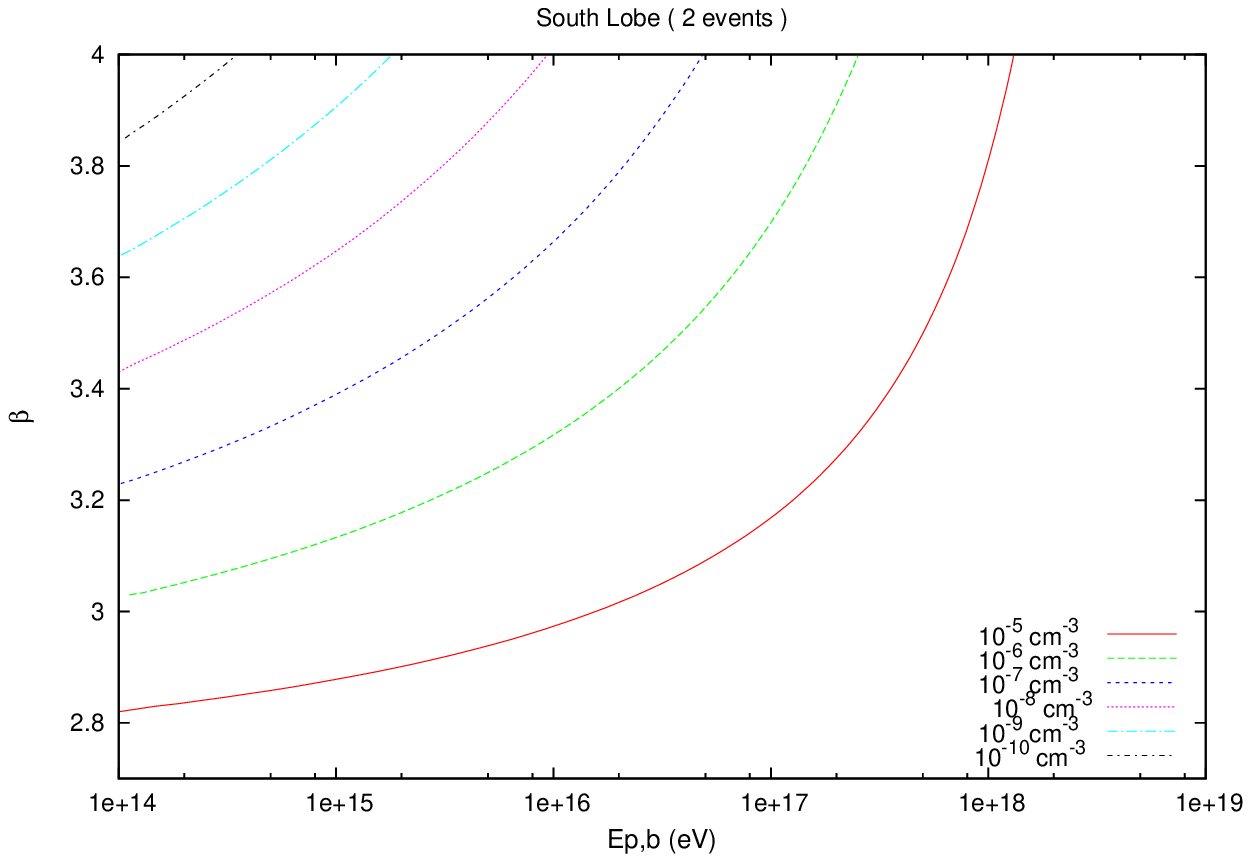}}
}
\caption{Contour plots  of higher spectral index ($\beta$) and  break energy ($E_{p,b}$) of the  broken power law of accelerated protons (eq. \ref{esppr2})  for which PAO would expect one (above) and two (below) events from the north (left) and south (right) lobes, when the number density of thermal particles is  $10^{-10} \,{\rm cm}^{-3} \leq  n_p  \leq 10^{-5}\, {\rm cm}^{-3}$. }
\label{c_plot}
\end{figure}

\begin{figure}
{\centering
\resizebox*{0.49\textwidth}{0.25\textheight}
{\includegraphics{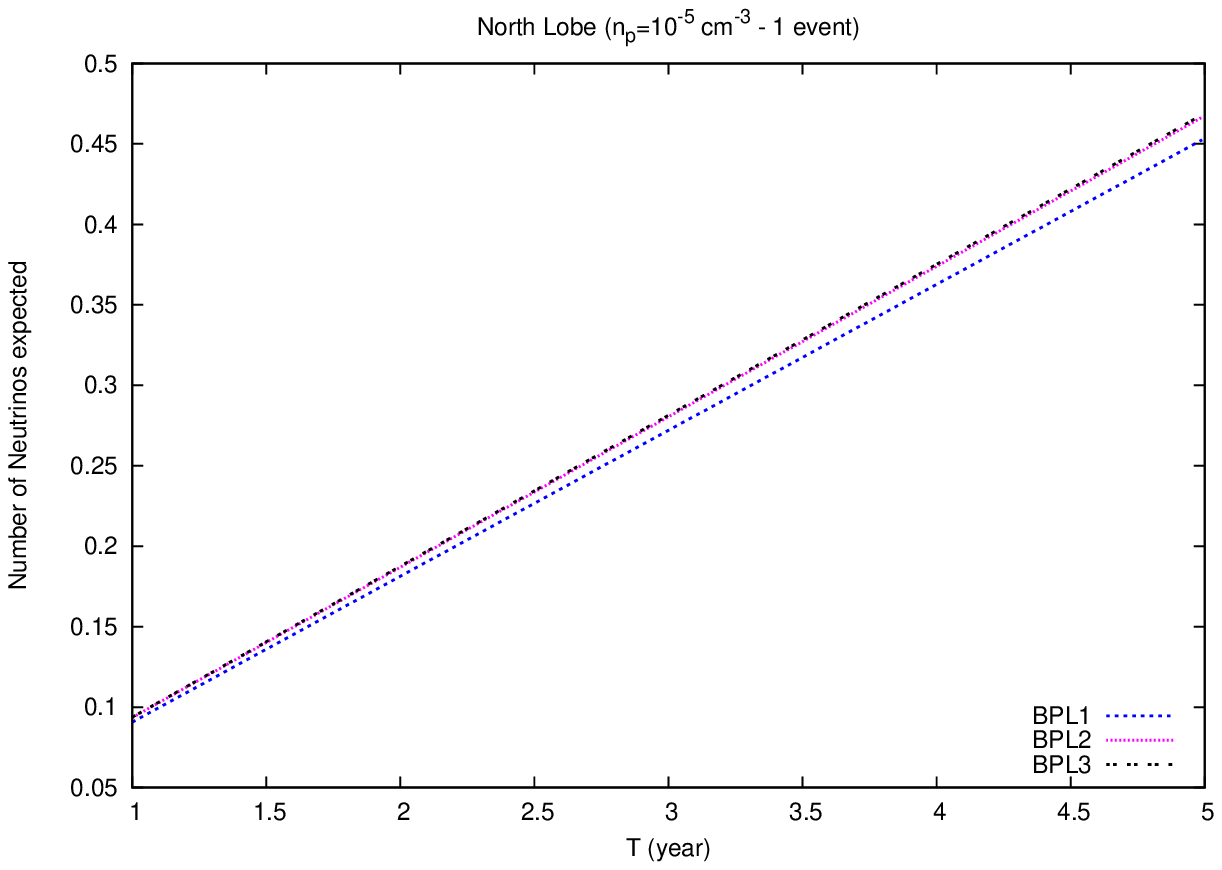}}
\resizebox*{0.49\textwidth}{0.25\textheight}
{\includegraphics{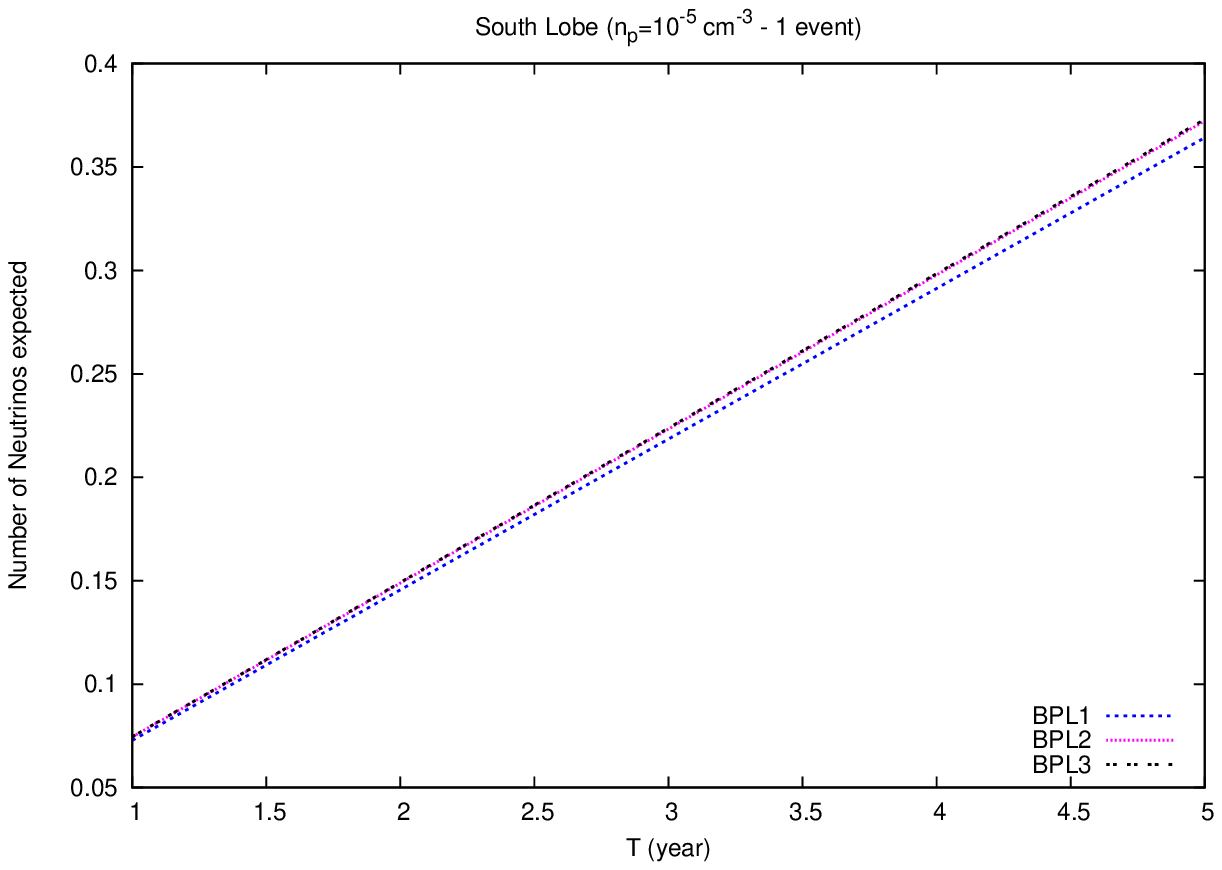}}
\resizebox*{0.49\textwidth}{0.25\textheight}
{\includegraphics{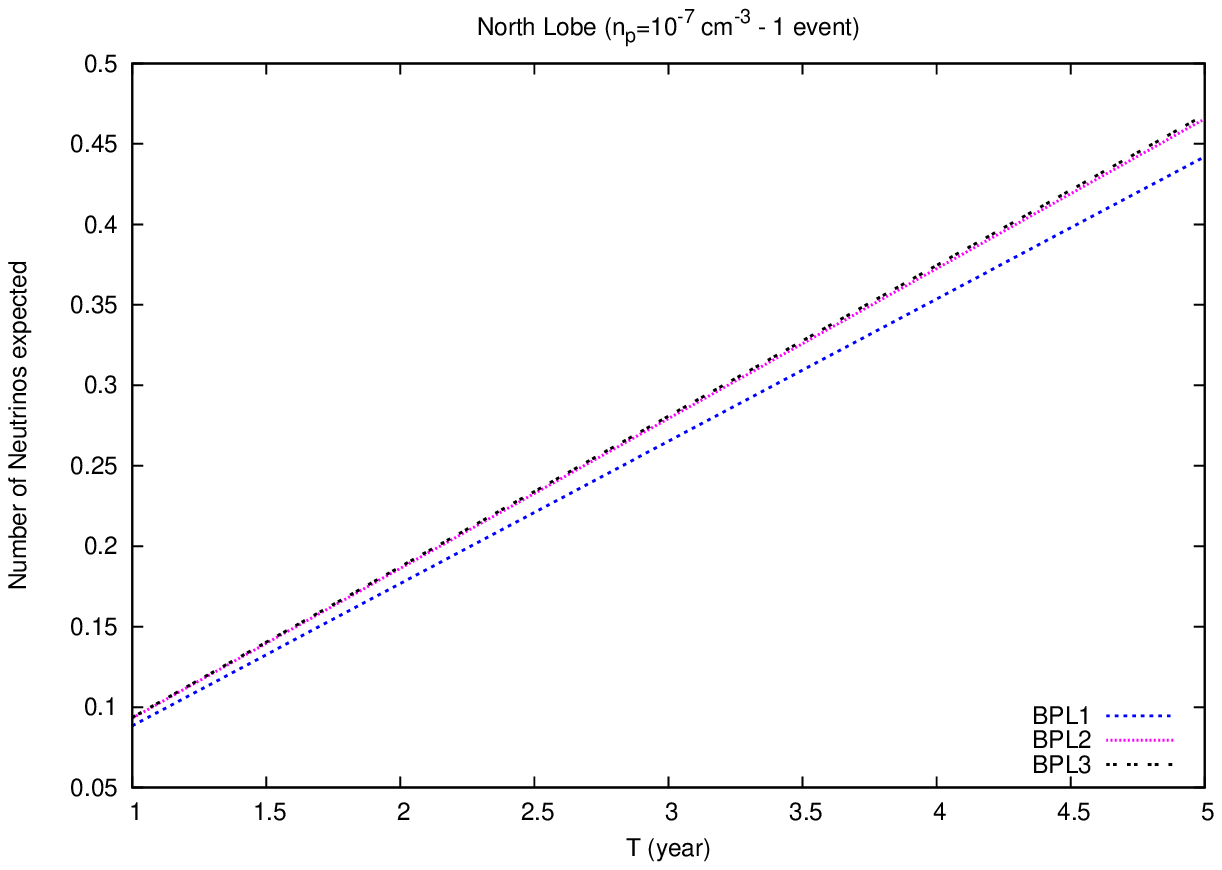}}
\resizebox*{0.49\textwidth}{0.25\textheight}
{\includegraphics{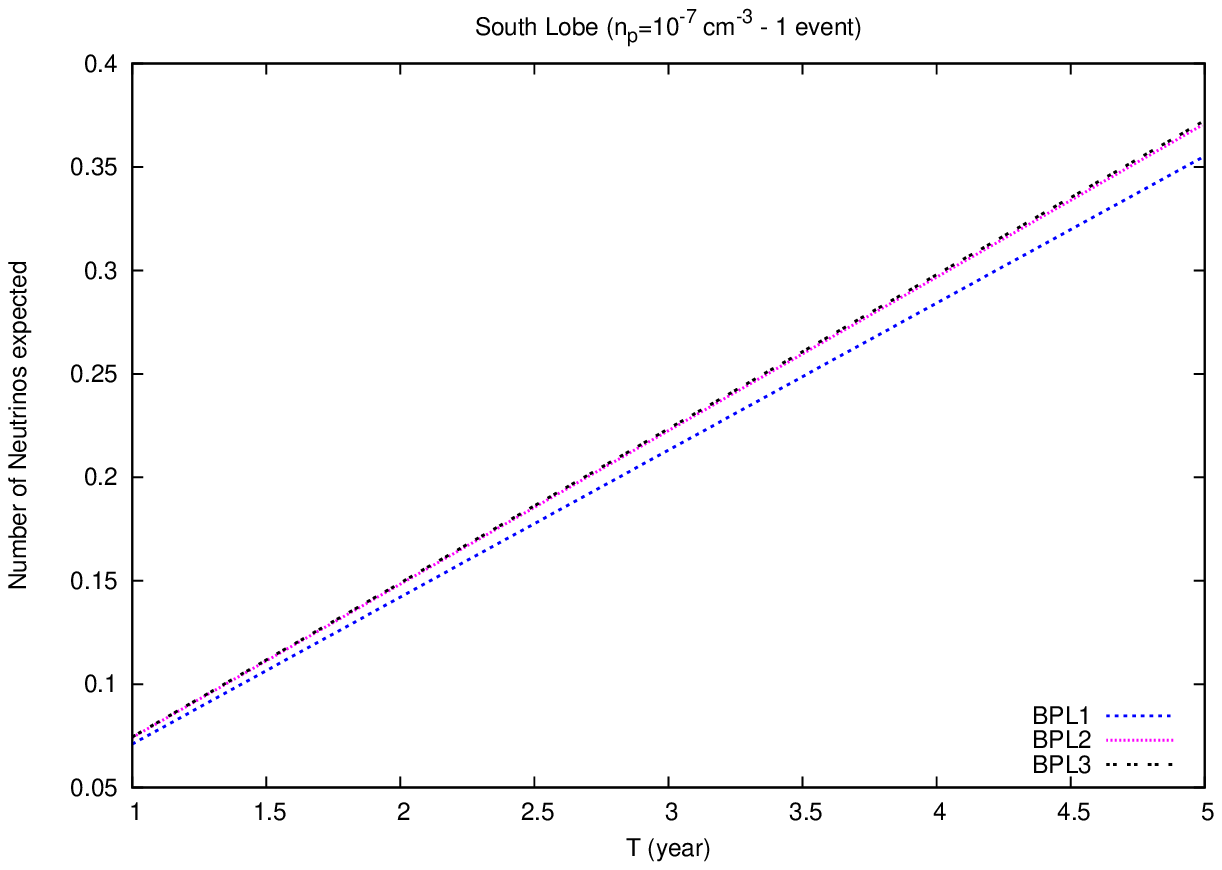}}
\resizebox*{0.49\textwidth}{0.25\textheight}
{\includegraphics{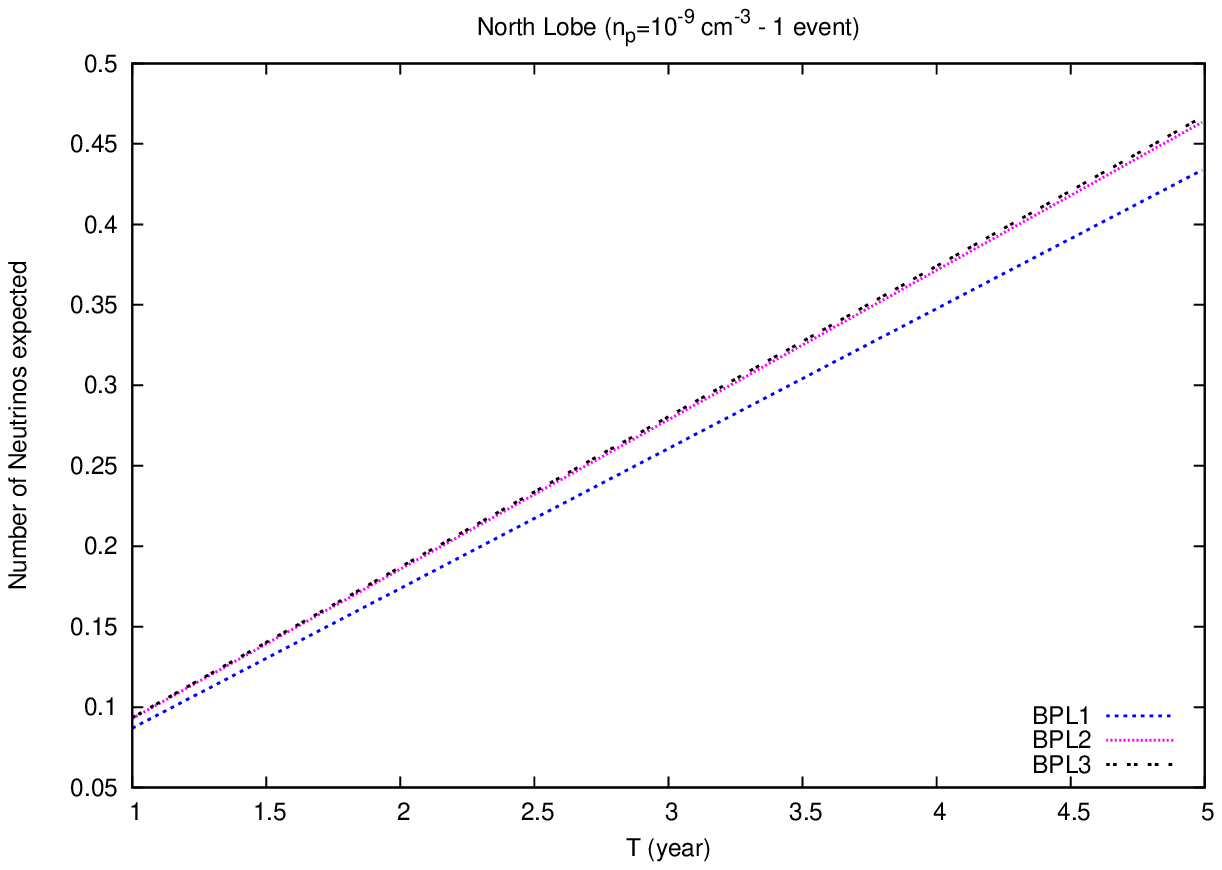}}
\resizebox*{0.49\textwidth}{0.25\textheight}
{\includegraphics{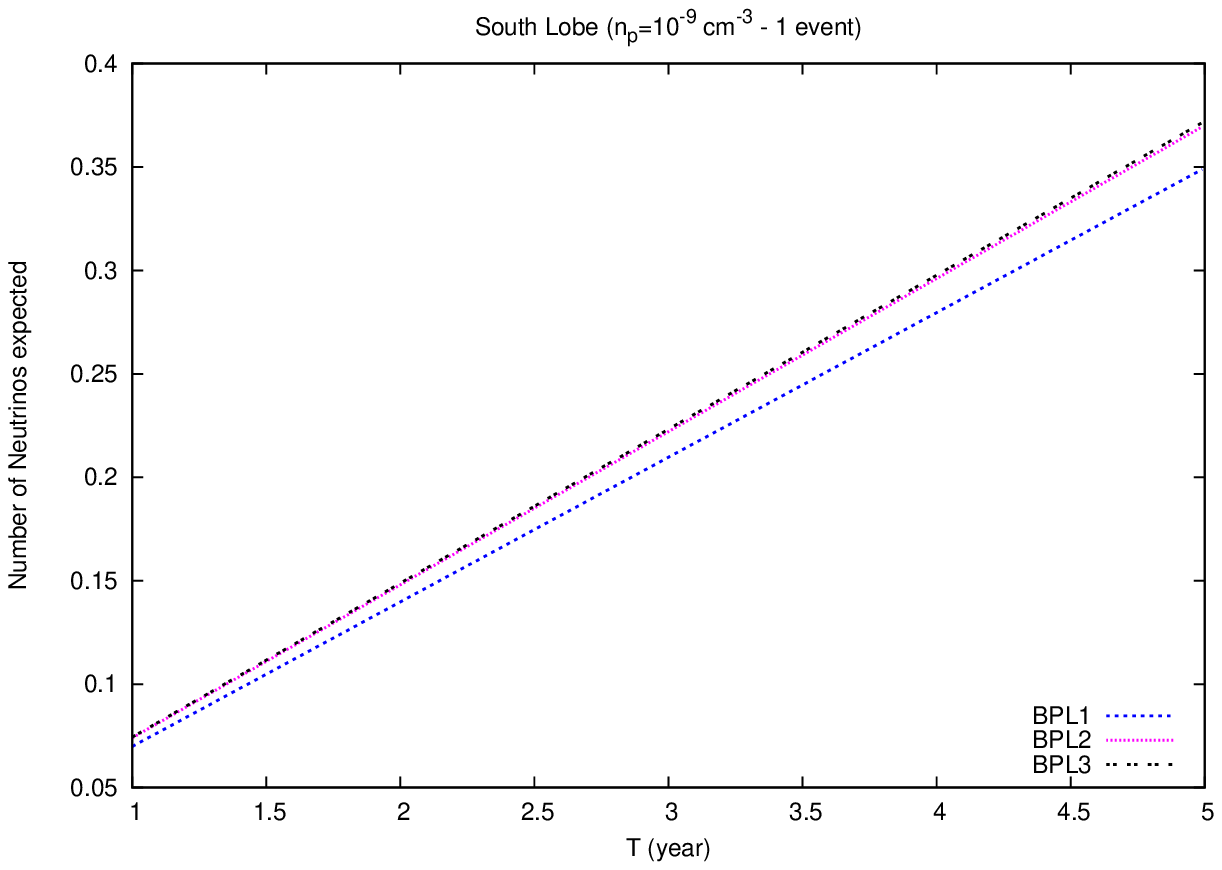}}
}
\caption{Number of neutrinos  expected on IceCube when the neutrino threshold energy ($E_{\nu,th}$) is 1 TeV. These neutrinos  are  generated by pp interactions and  normalized through  $\gamma$-ray from north (left) and south (right) lobes. These figures show the number of neutrinos produced taking into account the parameters $\beta$ and $E_{p,b}$ of the broken power law of accelerated protons for  which one UCHER would arrive in PAO.  We have used the values of parameters $\beta$ and $E_{\nu,b}$ given in tables B1 and B2.}
\label{n_neu1}
\end{figure}

\begin{figure}
{\centering
\resizebox*{0.49\textwidth}{0.25\textheight}
{\includegraphics{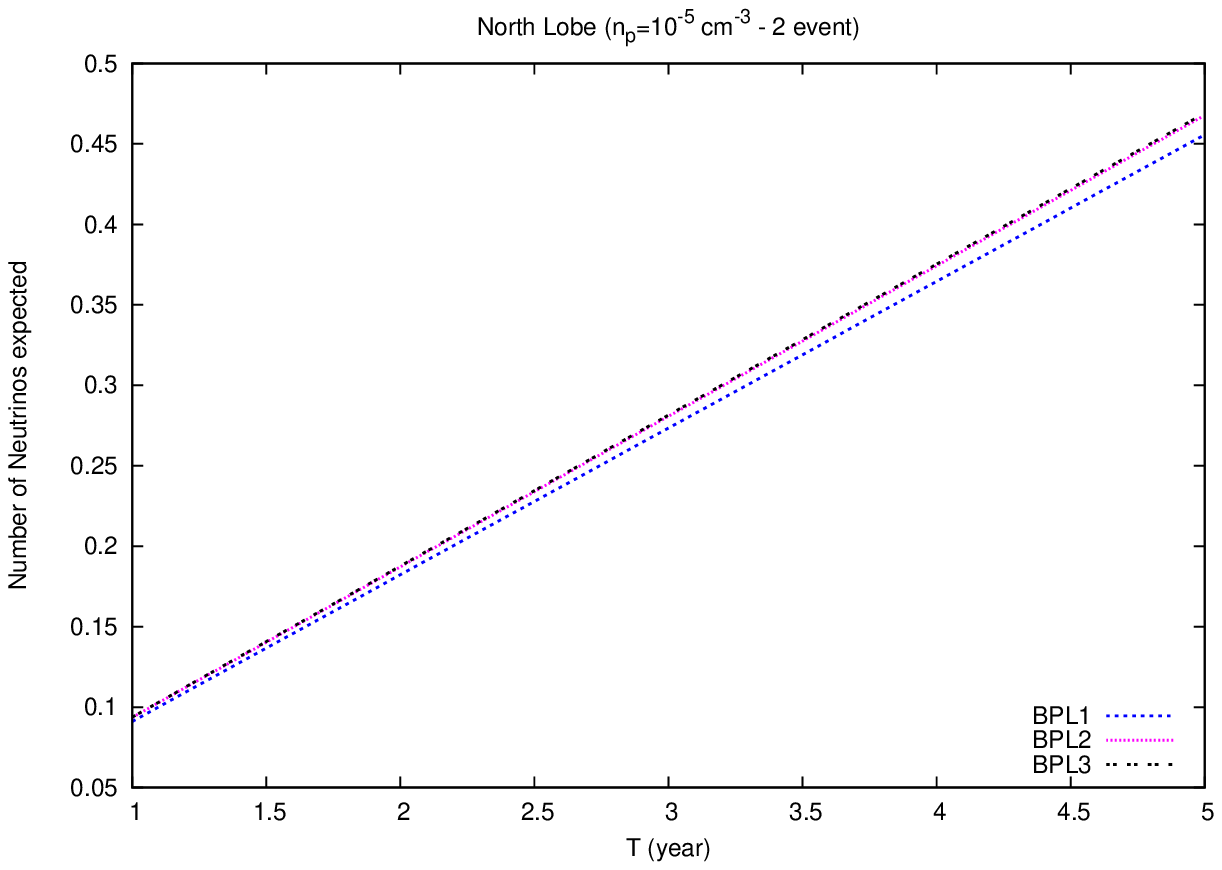}}
\resizebox*{0.49\textwidth}{0.25\textheight}
{\includegraphics{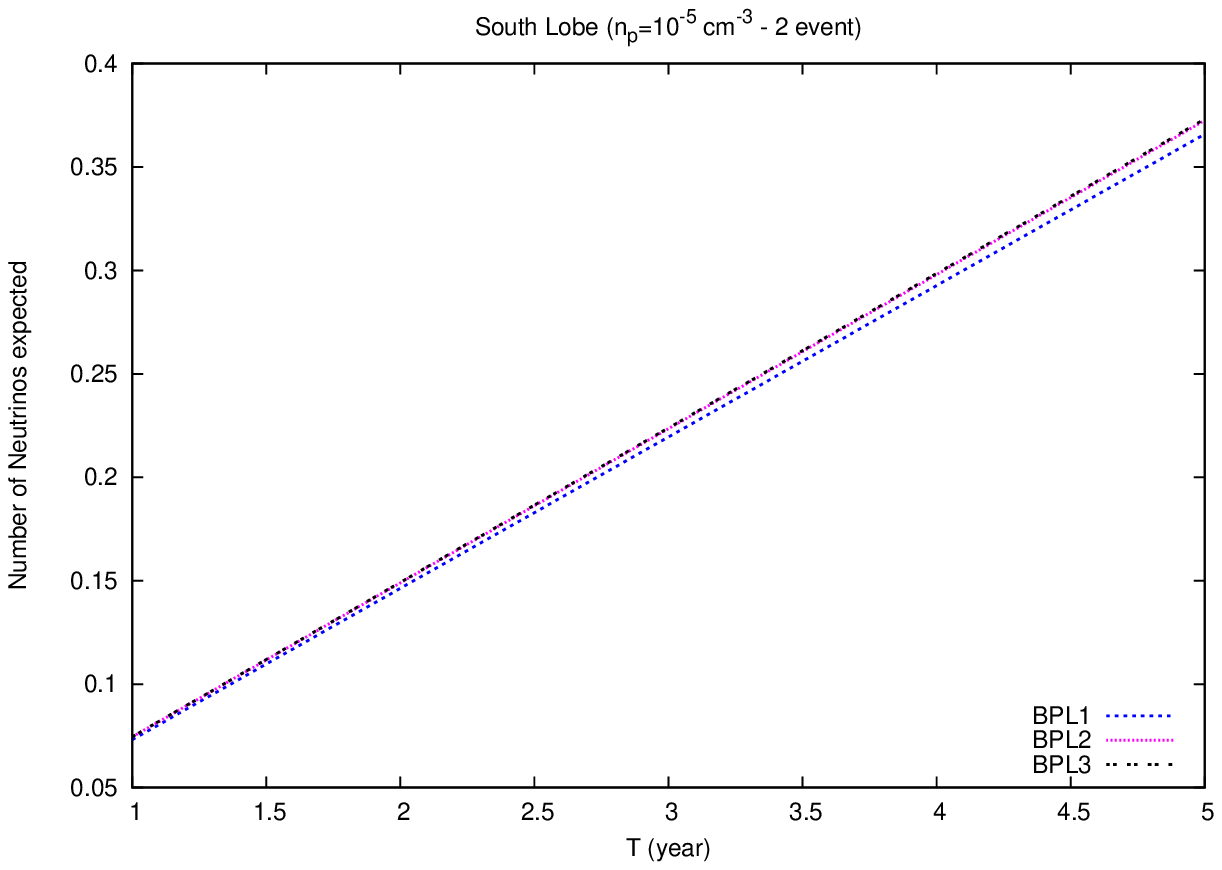}}
\resizebox*{0.49\textwidth}{0.25\textheight}
{\includegraphics{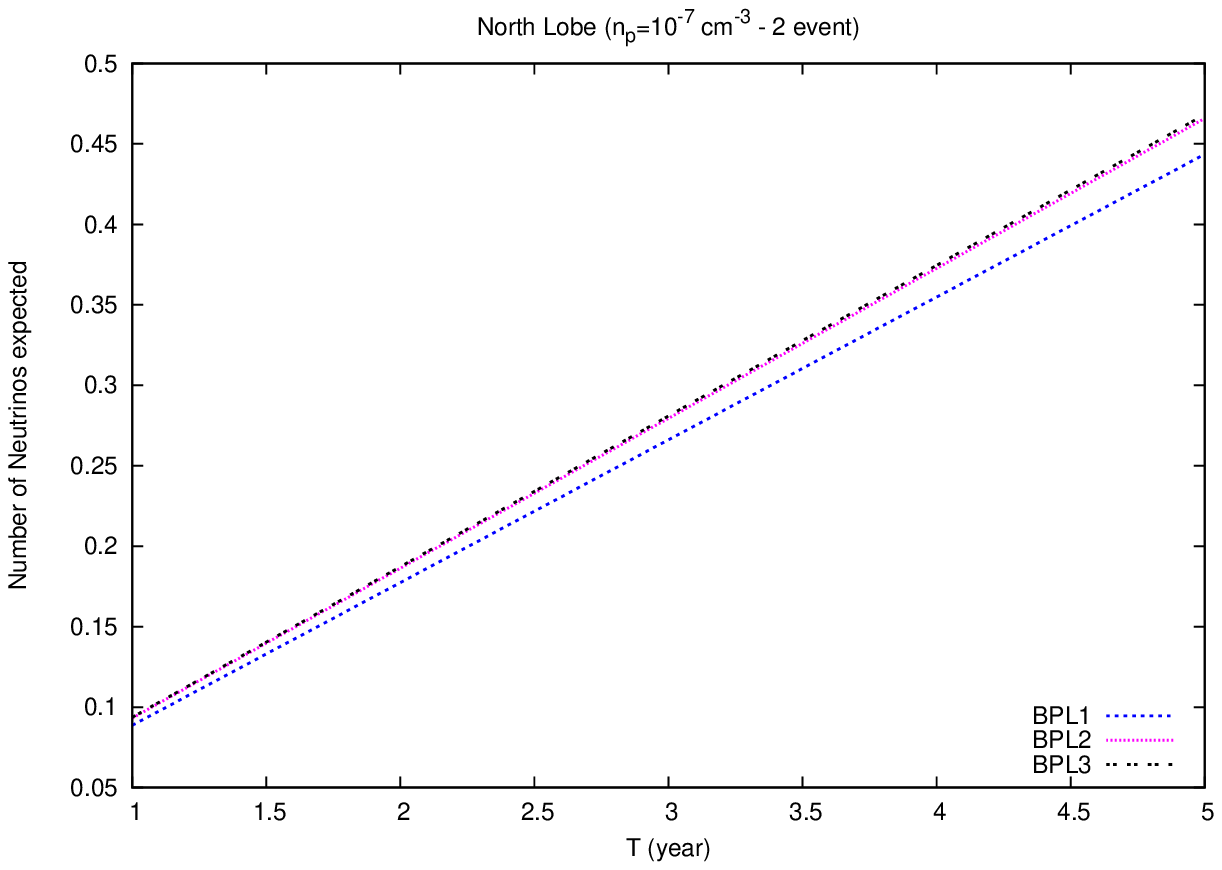}}
\resizebox*{0.49\textwidth}{0.25\textheight}
{\includegraphics{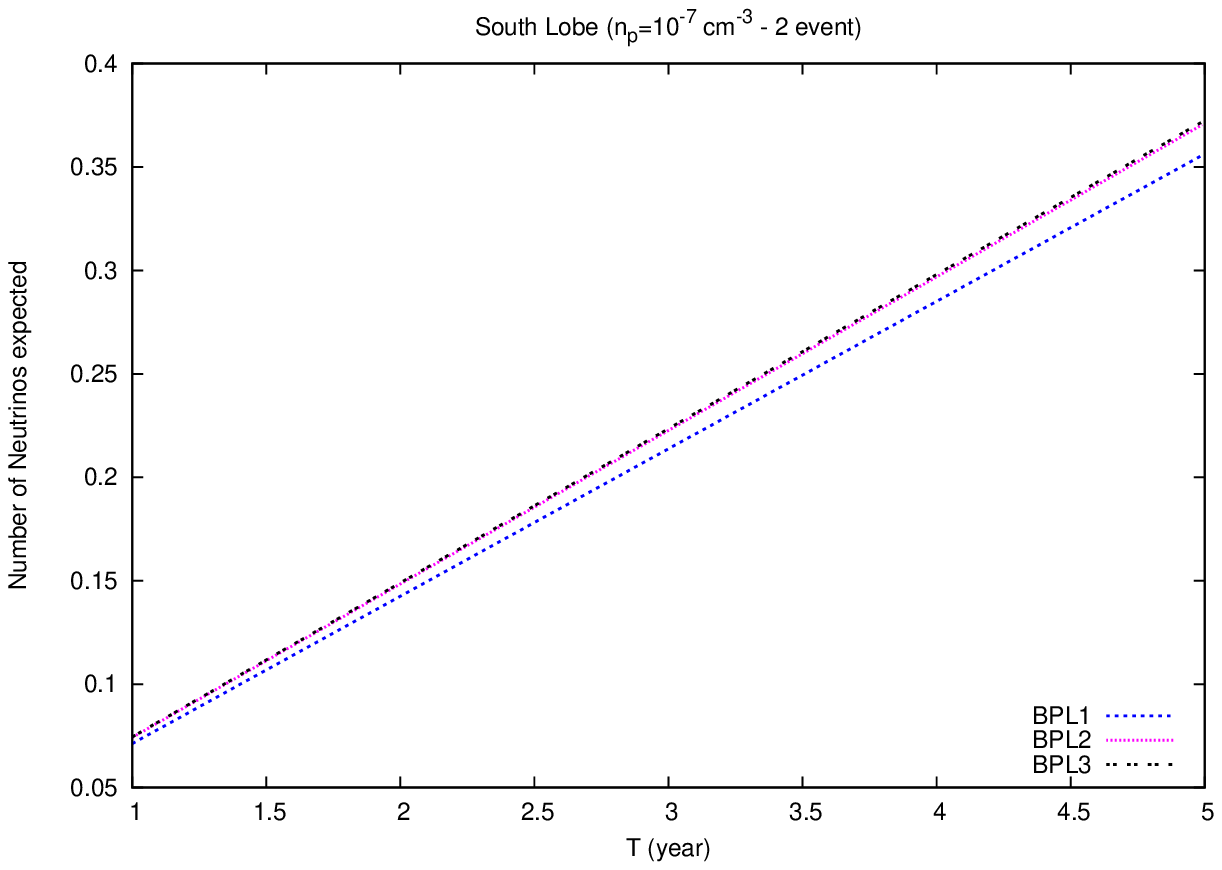}}
\resizebox*{0.49\textwidth}{0.25\textheight}
{\includegraphics{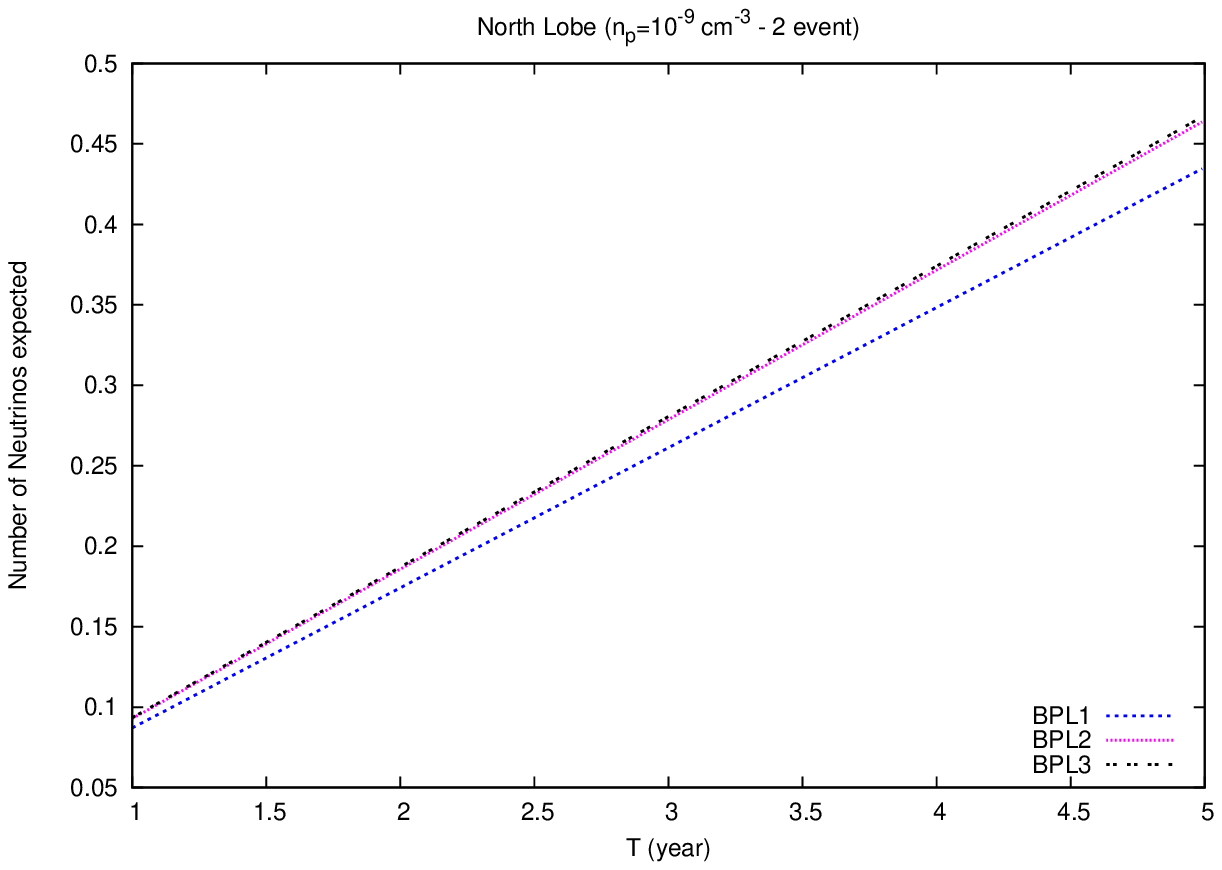}}
\resizebox*{0.49\textwidth}{0.25\textheight}
{\includegraphics{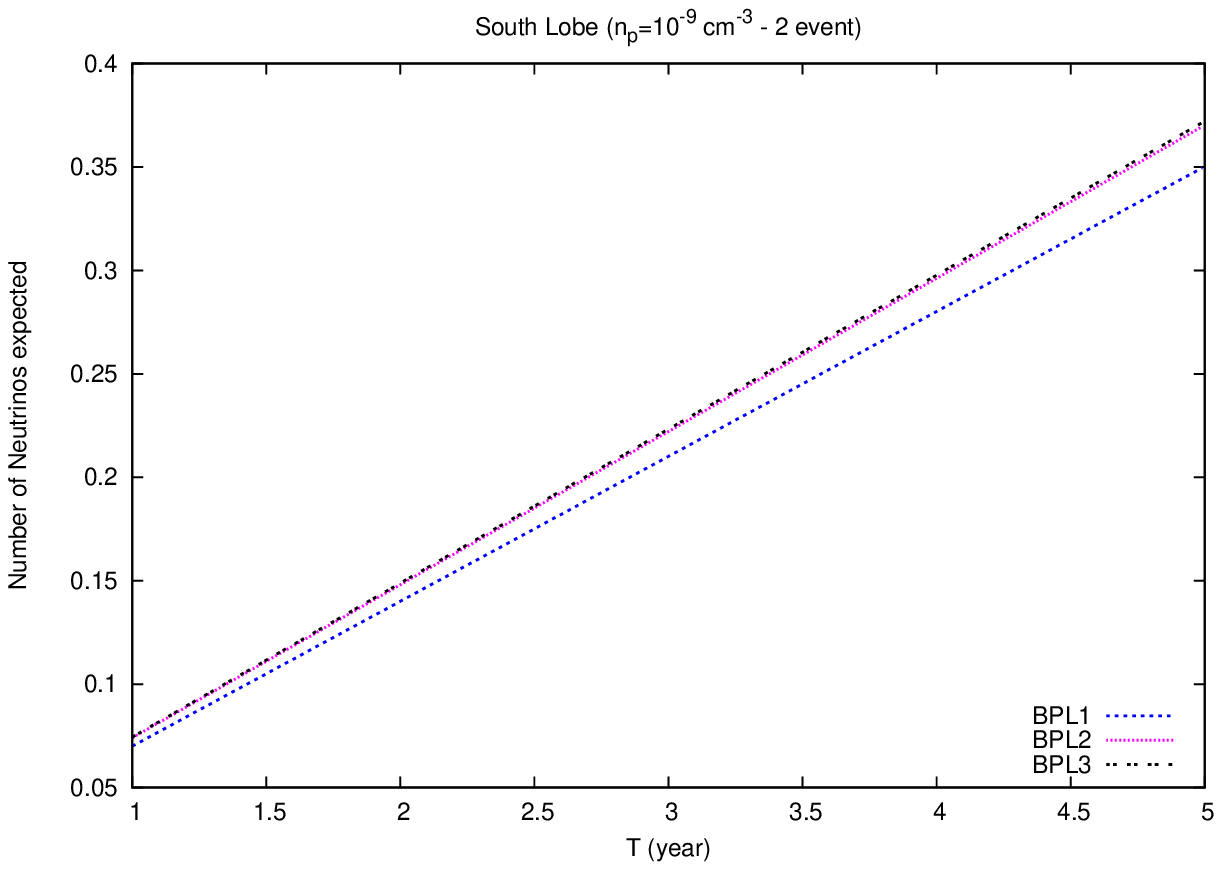}}
}
\caption{Number of neutrinos  expected on IceCube when the neutrino threshold energy ($E_{\nu,th}$) is 1 TeV. These neutrinos  are  generated by pp interactions and  normalized through  $\gamma$-ray from north (left) and south (right) lobes. These figures show the number of neutrinos produced taking into account the parameters $\beta$ and $E_{p,b}$ of the broken power law of accelerated protons for  which two UCHERs would arrive in PAO.  We have used the values of parameters $\beta$ and $E_{\nu,b}$ given in tables B1 and B2.}
\label{n_neu2}
\end{figure}

In the framework of emission processes,   we have done an exhaustive analysis to describe the photon spectrum of the lobes of Cen A.  In our emission model, firstly we have used synchrotron radiation to fit WMAP data and then estimated the values of magnetic fields, electron number densities as well as the age of the lobes, these values are calculated  assuming an equipartition between  the magnetic and electron energy density.  As shown in fig. \ref{fit_syn}, these quantities are plotted as a function of magnetic field for a wide range.  We estimate the values  (see section 4)   based on the choices of equipartition parameters; 4.3  and 1.8 for the north and south lobes, respectively  \citep{2010Sci...328..725A}, therefore the small difference  regarding the estimation of the age of north lobe given by \citep{2009MNRAS.393.1041H} comes of our election, although the age of south lobe  as well as the values of electron number densities in the lobes  represent  quite accurately \citet{2013A&A...558A..19W}.  Secondly,  we have fitted Fermi-LAT data with pp interactions in order to obtain the proton luminosity and then the non-thermal and thermal pressure  and the total energy in the lobes.   Although thermal particle densities in the range $10^{-10} \,{\rm cm}^{-3} \leq  n_p  \leq 10^{-4}\, {\rm cm}^{-3}$  describe successfully the $\gamma$-ray spectrum,  the density  $n_p  \sim 10^{-4}\, {\rm cm}^{-3}$ reproduces the  value of  proton luminosity $\sim 10^{43}\,erg\, s^{-1}$ which is more realistic in connection with the jet power  as well as  non-thermal  and thermal pressure   and  total energy  which have been estimated by using other methods \citep{2013ApJ...764..162O,2013ApJ...766...48S} (see table 2). Again one can see that the small differences come from the election of equipartition parameters and consequently the magnetic field.\\ 
 On the other hand, from the values of parameters found we have explored some acceleration mechanisms of UHECRs and showed that protons can be accelerated inside the lobes up to energies as high as $\sim 10^{20}$ eV, then we estimated the number of UHECRs expected in PAO,  supposing that  proton spectrum extends up to this energy range. We found that few events can be expected on Earth if and only if the thermal particles density is again $\sim 10^{-4}\, {\rm cm}^{-3}$. However, we investigated the conditions for which few events would arrive taking into account densities $\leq 10^{-5}\, {\rm cm}^{-3}$. We consider a broken power law for accelerating protons and  made contour plots of the spectrum parameters (fig. \ref{c_plot});  the higher spectral index ($\beta$) and the break energy ($E_{p,b}$) for which the expected number of UHECRs would be one event for each of the  lobes or two events for  one lobe.\\ 
On the other hand, correlating the $\gamma$-ray and neutrino fluxes  we have calculated the number of neutrinos expected  in IceCube.  Also we have considered a neutrino spectrum described by a simple and broken power law which are extended up to an energy range of IceCube. In both cases the number of neutrinos per year would be less than 0.938$\times 10^{-1}$ and 0.745$\times 10^{-1}$ for the north and south lobes, respectively, which is consistent with the non-detection of HE neutrinos by IceCube in the direction of Cen A \citep{2013arXiv1311.5238I, 2013arXiv1304.5356I}.\\
On the other hand and  as shown in fig. \ref{fit_ic}, our model is consistent to describe the $\gamma$-ray spectrum with  IC-scattering of CMB and EBL firstly proposed and discussed by \citet{2010Sci...328..725A, 2009MNRAS.393.1041H}.   Additionally, we have briefly introduced  p$\gamma$ interactions and showed that they did not contribute  to the $\gamma$-ray spectrum.\\
\acknowledgments
We thank the referee  for a critical reading of the paper and valuable suggestions.  We also thank Charles Dermer, Bin Zhang, Francis Halzen,  Ignacio Taboada, William Lee and Antonio Marinelli for useful discussions.  NF gratefully acknowledges a Luc Binette-Fundaci\'on UNAM Posdoctoral Fellowship.

%\bibliography{Bib_cena}
% (uses file "plain.bst")
%\mbox{}
%\addcontentsline{toc}{chapter}{Bibliography}

\clearpage

\clearpage
\appendix 
\section{Chi-square minimization}
Firstly, from pp interaction model  (eq. \ref{pp}),  we fit the $\gamma$-ray spectrum  using two parameters, the proportionality constant of pp spectrum $A_{pp,\gamma}$ (eq. \ref{App}) and the spectral index $\alpha$, as follow.
\begin{equation}
\label{ppf}
\left(E^{2}_\gamma\, \frac{dN_\gamma}{dE_\gamma}\right)^{obs}_{\pi^0}=[0]\, \left(\frac{E^{obs}_{\gamma,\pi^0}}{{\rm GeV}}\right)^{2-[1]},
\end{equation}
After fitting we obtained the values
\begin{center}\renewcommand{\arraystretch}{0.6}\addtolength{\tabcolsep}{-1pt}
\begin{tabular}{ l c c c c}
  \hline \hline
 Lobes & \scriptsize{} & & \scriptsize{North} & \scriptsize{South} \\
 \hline \hline
 \scriptsize{} &\scriptsize{Parameter} & \scriptsize{Symbol} & \scriptsize{Value} & \scriptsize{Value} \\
 \hline
\hline
\scriptsize{Proportionality constant} ($10^{-12}\,{\rm erg/cm^2/s}$)  &\scriptsize{[0]}                    & \scriptsize{$ A_{pp,\gamma} $}  &  \scriptsize{ $5.10\pm 0.96$} &  \scriptsize{ $8.07\pm 1.58$}\\
\scriptsize{Spectral index}                                                                 &\scriptsize{[1]}                    & \scriptsize{$\alpha$}  & \scriptsize{2.519$\pm$ 0.225} & \scriptsize{2.598$\pm$ 0.254} \\
\scriptsize{Chi-square/NDF}                                                              & & \scriptsize{$ \chi^2/{\rm NDF}$}  &  \scriptsize{ $2.748/4.0$} &  \scriptsize{ $0.555/3.0$}\\
\hline
\end{tabular}
\end{center}

\begin{center}
\scriptsize{\textbf{Table A1.  The best fit of the set of pp interaction parameters obtained  after fitting the $\gamma$-ray spectrum  of north and south lobes.}}\\
\end{center}
Secondly, from synchrotron emission model  (eq.  \ref{espsyn})  and  the value of $\alpha$ (table 1), we fit  the peak at  radio wavelength using  three parameters,  the proportionality constant of synchrotron $A_{syn,\gamma}$ (eq. \ref{Asyn}), the characteristic and cut-off photon energies $\epsilon^{obs}_{\gamma,m}$   and   $\epsilon^{obs}_{\gamma,c}$ (eq. \ref{synrad}), respectively  as follow 
{\small
\begin{equation}
\label{espsynf}
\epsilon^2_\gamma N_\gamma(\epsilon_\gamma) = [0]
\cases {
(\frac{\epsilon_\gamma}{[2]})^{4/3}    &  $\epsilon^{obs}_\gamma < [2]$,\cr
 (\frac{\epsilon_\gamma}{[2]})^{-(\alpha-3)/2}  &  $[2] < \epsilon^{obs}_\gamma < [1]$,\cr
(\frac{[1]}{[2]})^{-(\alpha-3)/2}    (\frac{\epsilon_\gamma}{[1]})^{-(\alpha-2)/2},           &  $[1] < \epsilon^{obs}_\gamma  $\cr
}
\end{equation}
\small}
The values of the parameters obtained after fitting
\begin{center}\renewcommand{\arraystretch}{0.6}\addtolength{\tabcolsep}{-1pt}
\begin{tabular}{ l c c c c}
  \hline \hline
 Lobes & \scriptsize{} & \scriptsize{North} & \scriptsize{South} \\
 \hline \hline
 \scriptsize{} & \scriptsize{Parameter} &\scriptsize{Symbol} & \scriptsize{Value} & \scriptsize{Value} \\
 \hline
\hline
\scriptsize{Proportionality constant} ($10^{-12}\,{\rm erg/cm^2/s}$) &\scriptsize{[0]}  & \scriptsize{$ A_{syn,\gamma} $}  &  \scriptsize{ $1.08\pm 0.30$} &  \scriptsize{ $4.51\pm 0.64$}\\
\scriptsize{Cut-off photon energy} ($10^{-5}\,{\rm eV}$)                   &\scriptsize{[1]}  & \scriptsize{$ \epsilon^{obs}_{\gamma,c} $}  &  \scriptsize{ $5.703\pm 0.557$} &  \scriptsize{ $4.01\pm 0.71$}\\
\scriptsize{Characteristic photon energy}  ($10^{-6}\,{\rm eV}$)      &\scriptsize{[2]}   & \scriptsize{$\epsilon^{obs}_{\gamma,m}$}  & \scriptsize{2.63$\pm$ 0.03} & \scriptsize{3.48$\pm$ 1.70} \\
\scriptsize{Chi-square/NDF}                                                             & & \scriptsize{$ \chi^2/{\rm NDF}$}  &  \scriptsize{ $0.8616/2.0$} &  \scriptsize{ $1.290/4.0$}\\
 \hline
\end{tabular}
\end{center}
\begin{center}
\scriptsize{\textbf{Table A2.  The best fit of the set of synchrotron parameters obtained  after fitting  spectrum at low energies  of north and south lobes.}}\\
\end{center}
\section{Neutrinos broken power law}

\begin{center}\renewcommand{\arraystretch}{0.85}\addtolength{\tabcolsep}{-1pt}
\begin{tabular}{c c c c c c c c}
  \hline \hline
&  &  one &event &\hspace{0.5 cm} &&two &events  \\
 \hline 
&\scriptsize{$\beta$} &  \scriptsize{$E_{\nu,b} $ (eV)} & \scriptsize{$N_{ev}/T (year)^{-1}$} &         &\scriptsize{$\beta$} &  \scriptsize{$E_{\nu,b} $ (eV)} & \scriptsize{$N_{ev}/T (year)^{-1}$} \\
 \hline
 \scriptsize{$n_p=10^{-5}\,{\rm cm}^{-3} $}   & & & & & & &\\ 
 \hline
\scriptsize{BPL1}&\scriptsize{2.87}  &\scriptsize{$0.5\times 10^{13}$} & \scriptsize{$9.07\times 10^{-2}$} & & \scriptsize{2.81} & \scriptsize{$0.5\times 10^{13}$} &\scriptsize{$9.11\times 10^{-2}$} \\
\scriptsize{BPL2}&\scriptsize{2.92}  &\scriptsize{$2.5\times 10^{13}$} & \scriptsize{$9.35\times 10^{-2}$}& &  \scriptsize{2.85} & \scriptsize{$2.5\times 10^{13}$} &\scriptsize{$9.36\times 10^{-2}$} \\
\scriptsize{BPL3}&\scriptsize{2.95}  &\scriptsize{$0.5\times 10^{14}$} & \scriptsize{$9.38\times 10^{-2}$}& &  \scriptsize{2.88} & \scriptsize{$0.5\times 10^{14}$} &\scriptsize{$9.38\times 10^{-2}$} \\
 \hline
 \scriptsize{$n_p=10^{-6}\,{\rm cm}^{-3} $}   & & & & & & &\\ 
 \hline
\scriptsize{BPL1}&\scriptsize{3.06}  &\scriptsize{$0.5\times 10^{13}$} & \scriptsize{$8.94\times 10^{-2}$} & & \scriptsize{3.00} & \scriptsize{$0.5\times 10^{13}$} &\scriptsize{$8.98\times 10^{-2}$} \\
\scriptsize{BPL2}&\scriptsize{3.14}  &\scriptsize{$2.5\times 10^{13}$} & \scriptsize{$9.33\times 10^{-2}$}& &  \scriptsize{3.07} & \scriptsize{$2.5\times 10^{13}$} &\scriptsize{$9.33\times 10^{-2}$} \\
\scriptsize{BPL3}&\scriptsize{3.18}  &\scriptsize{$0.5\times 10^{14}$} & \scriptsize{$9.37\times 10^{-2}$}& &  \scriptsize{3.11} & \scriptsize{$0.5\times 10^{14}$} &\scriptsize{$9.37\times 10^{-2}$} \\
 \hline
 \scriptsize{$n_p=10^{-7}\,{\rm cm}^{-3} $}   & & & & & & &\\ 
 \hline
\scriptsize{BPL1}&\scriptsize{3.25}  &\scriptsize{$0.5\times 10^{13}$} & \scriptsize{$8.84\times 10^{-2}$} & & \scriptsize{3.19} & \scriptsize{$0.5\times 10^{13}$} &\scriptsize{$8.87\times 10^{-2}$} \\
\scriptsize{BPL2}&\scriptsize{3.36}  &\scriptsize{$2.5\times 10^{13}$} & \scriptsize{$9.31\times 10^{-2}$}& &  \scriptsize{3.29} & \scriptsize{$2.5\times 10^{13}$} &\scriptsize{$9.32\times 10^{-2}$} \\
\scriptsize{BPL3}&\scriptsize{3.42}  &\scriptsize{$0.5\times 10^{14}$} & \scriptsize{$9.37\times 10^{-2}$}& &  \scriptsize{3.35} & \scriptsize{$0.5\times 10^{14}$} &\scriptsize{$9.36\times 10^{-2}$} \\
 \hline
 \scriptsize{$n_p=10^{-8}\,{\rm cm}^{-3} $}   & & & & & & &\\ 
 \hline
\scriptsize{BPL1}&\scriptsize{3.44}  &\scriptsize{$0.5\times 10^{13}$} & \scriptsize{$8.76\times 10^{-2}$} & & \scriptsize{3.38} & \scriptsize{$0.5\times 10^{13}$} &\scriptsize{$8.78\times 10^{-2}$} \\
\scriptsize{BPL2}&\scriptsize{3.58}  &\scriptsize{$2.5\times 10^{13}$} & \scriptsize{$9.30\times 10^{-2}$}& &  \scriptsize{3.51} & \scriptsize{$2.5\times 10^{13}$} &\scriptsize{$9.30\times 10^{-2}$} \\
\scriptsize{BPL3}&\scriptsize{3.66}  &\scriptsize{$0.5\times 10^{14}$} & \scriptsize{$9.36\times 10^{-2}$}& &  \scriptsize{3.59} & \scriptsize{$0.5\times 10^{14}$} &\scriptsize{$9.35\times 10^{-2}$} \\
 \hline
 \scriptsize{$n_p=10^{-9}\,{\rm cm}^{-3} $}   & & & & & & &\\ 
 \hline
\scriptsize{BPL1}&\scriptsize{3.63}  &\scriptsize{$0.5\times 10^{13}$} & \scriptsize{$8.69\times 10^{-2}$} & & \scriptsize{3.58} & \scriptsize{$0.5\times 10^{13}$} &\scriptsize{$8.71\times 10^{-2}$} \\
\scriptsize{BPL2}&\scriptsize{3.80}  &\scriptsize{$2.5\times 10^{13}$} & \scriptsize{$9.29\times 10^{-2}$}& &  \scriptsize{3.74} & \scriptsize{$2.5\times 10^{13}$} &\scriptsize{$9.29\times 10^{-2}$} \\
\scriptsize{BPL3}&\scriptsize{3.90}  &\scriptsize{$0.5\times 10^{14}$} & \scriptsize{$9.35\times 10^{-2}$}& &  \scriptsize{3.83} & \scriptsize{$0.5\times 10^{14}$} &\scriptsize{$9.35\times 10^{-2}$} \\
\hline
 \scriptsize{$n_p=10^{-10}\,{\rm cm}^{-3} $}   & & & & & & &\\ 
 \hline
\scriptsize{BPL1}&\scriptsize{3.83}  &\scriptsize{$0.5\times 10^{13}$} & \scriptsize{$8.63\times 10^{-2}$} & & \scriptsize{3.77} & \scriptsize{$0.5\times 10^{13}$} &\scriptsize{$8.65\times 10^{-2}$} \\
\scriptsize{BPL2}&\scriptsize{4.03}  &\scriptsize{$2.5\times 10^{13}$} & \scriptsize{$9.28\times 10^{-2}$}& &  \scriptsize{3.96} & \scriptsize{$2.5\times 10^{13}$} &\scriptsize{$9.28\times 10^{-2}$} \\
\scriptsize{BPL3}&\scriptsize{4.14}  &\scriptsize{$0.5\times 10^{14}$} & \scriptsize{$9.35\times 10^{-2}$}& &  \scriptsize{4.07} & \scriptsize{$0.5\times 10^{14}$} &\scriptsize{$9.35\times 10^{-2}$} \\
 \hline
\end{tabular}
\end{center}

\begin{center}
\scriptsize{\textbf{Table B1. Number of neutrinos above 1 TeV expected per year from the north lobe (see fig. 3  of the left).  Here we show the parameters ($\beta$ and $E_{\nu,b}$)  of neutrino spectrum (eq. \ref{espneu2}) for three broken power laws (BPLs) when one and two UCHERs are expected in PAO.}}\\
\end{center}

\begin{center}\renewcommand{\arraystretch}{0.85}\addtolength{\tabcolsep}{-1pt}
\begin{tabular}{c c c c c c c c}
  \hline \hline
&  &  one &event &\hspace{0.5 cm} &&two &events  \\
 \hline 
&\scriptsize{$\beta$} &  \scriptsize{$E_{\nu,b} $ (eV)} & \scriptsize{$N_{ev}/T (year)^{-1}$} &         &\scriptsize{$\beta$} &  \scriptsize{$E_{\nu,b} $ (eV)} & \scriptsize{$N_{ev}/T (year)^{-1}$} \\
 \hline
 \scriptsize{$n_p=10^{-5}\,{\rm cm}^{-3} $}   & & & & & & &\\ 
 \hline
\scriptsize{BPL1}&\scriptsize{2.88}  &\scriptsize{$0.5\times 10^{13}$} & \scriptsize{$7.28\times 10^{-2}$} & & \scriptsize{2.82} & \scriptsize{$0.5\times 10^{13}$} &\scriptsize{$7.32\times 10^{-2}$} \\
\scriptsize{BPL2}&\scriptsize{2.93}  &\scriptsize{$2.5\times 10^{13}$} & \scriptsize{$7.44\times 10^{-2}$}& &  \scriptsize{2.86} & \scriptsize{$2.5\times 10^{13}$} &\scriptsize{$7.45\times 10^{-2}$} \\
\scriptsize{BPL3}&\scriptsize{2.95}  &\scriptsize{$0.5\times 10^{14}$} & \scriptsize{$7.46\times 10^{-2}$}& &  \scriptsize{2.88} & \scriptsize{$0.5\times 10^{14}$} &\scriptsize{$7.46\times 10^{-2}$} \\
 \hline
 \scriptsize{$n_p=10^{-6}\,{\rm cm}^{-3} $}   & & & & & & &\\ 
 \hline
\scriptsize{BPL1}&\scriptsize{3.09}  &\scriptsize{$0.5\times 10^{13}$} & \scriptsize{$7.18\times 10^{-2}$} & & \scriptsize{3.03} & \scriptsize{$0.5\times 10^{13}$} &\scriptsize{$7.21\times 10^{-2}$} \\
\scriptsize{BPL2}&\scriptsize{3.17}  &\scriptsize{$2.5\times 10^{13}$} & \scriptsize{$7.43\times 10^{-2}$}& &  \scriptsize{3.09} & \scriptsize{$2.5\times 10^{13}$} &\scriptsize{$7.43\times 10^{-2}$} \\
\scriptsize{BPL3}&\scriptsize{3.21}  &\scriptsize{$0.5\times 10^{14}$} & \scriptsize{$7.45\times 10^{-2}$}& &  \scriptsize{3.13} & \scriptsize{$0.5\times 10^{14}$} &\scriptsize{$7.46\times 10^{-2}$} \\
 \hline
 \scriptsize{$n_p=10^{-7}\,{\rm cm}^{-3} $}   & & & & & & &\\ 
 \hline
\scriptsize{BPL1}&\scriptsize{3.29}  &\scriptsize{$0.5\times 10^{13}$} & \scriptsize{$7.10\times 10^{-2}$} & & \scriptsize{3.23} & \scriptsize{$0.5\times 10^{13}$} &\scriptsize{$7.13\times 10^{-2}$} \\
\scriptsize{BPL2}&\scriptsize{3.40}  &\scriptsize{$2.5\times 10^{13}$} & \scriptsize{$7.42\times 10^{-2}$}& &  \scriptsize{3.33} & \scriptsize{$2.5\times 10^{13}$} &\scriptsize{$7.42\times 10^{-2}$} \\
\scriptsize{BPL3}&\scriptsize{3.47}  &\scriptsize{$0.5\times 10^{14}$} & \scriptsize{$7.45\times 10^{-2}$}& &  \scriptsize{3.39} & \scriptsize{$0.5\times 10^{14}$} &\scriptsize{$7.45\times 10^{-2}$} \\
 \hline
 \scriptsize{$n_p=10^{-8}\,{\rm cm}^{-3} $}   & & & & & & &\\ 
 \hline
\scriptsize{BPL1}&\scriptsize{3.49}  &\scriptsize{$0.5\times 10^{13}$} & \scriptsize{$7.04\times 10^{-2}$} & & \scriptsize{3.43} & \scriptsize{$0.5\times 10^{13}$} &\scriptsize{$7.06\times 10^{-2}$} \\
\scriptsize{BPL2}&\scriptsize{3.64}  &\scriptsize{$2.5\times 10^{13}$} & \scriptsize{$7.41\times 10^{-2}$}& &  \scriptsize{3.57} & \scriptsize{$2.5\times 10^{13}$} &\scriptsize{$7.41\times 10^{-2}$} \\
\scriptsize{BPL3}&\scriptsize{3.72}  &\scriptsize{$0.5\times 10^{14}$} & \scriptsize{$7.45\times 10^{-2}$}& &  \scriptsize{3.65} & \scriptsize{$0.5\times 10^{14}$} &\scriptsize{$9.45\times 10^{-2}$} \\
 \hline
\scriptsize{$n_p=10^{-9}\,{\rm cm}^{-3} $}   & & & & & & &\\ 
 \hline
\scriptsize{BPL1}&\scriptsize{3.70}  &\scriptsize{$0.5\times 10^{13}$} & \scriptsize{$6.99\times 10^{-2}$} & & \scriptsize{3.64} & \scriptsize{$0.5\times 10^{13}$} &\scriptsize{$7.00\times 10^{-2}$} \\
\scriptsize{BPL2}&\scriptsize{3.88}  &\scriptsize{$2.5\times 10^{13}$} & \scriptsize{$7.40\times 10^{-2}$}& &  \scriptsize{3.81} & \scriptsize{$2.5\times 10^{13}$} &\scriptsize{$7.40\times 10^{-2}$} \\
\scriptsize{BPL3}&\scriptsize{3.98}  &\scriptsize{$0.5\times 10^{14}$} & \scriptsize{$7.44\times 10^{-2}$}& &  \scriptsize{3.91} & \scriptsize{$0.5\times 10^{14}$} &\scriptsize{$7.44\times 10^{-2}$} \\
\hline
 \scriptsize{$n_p=10^{-10}\,{\rm cm}^{-3} $}   & & & & & & &\\ 
 \hline
\scriptsize{BPL1}&\scriptsize{3.90}  &\scriptsize{$0.5\times 10^{13}$} & \scriptsize{$6.95\times 10^{-2}$} & & \scriptsize{3.85} & \scriptsize{$0.5\times 10^{13}$} &\scriptsize{$6.96\times 10^{-2}$} \\
\scriptsize{BPL2}&\scriptsize{4.12}  &\scriptsize{$2.5\times 10^{13}$} & \scriptsize{$7.40\times 10^{-2}$}& &  \scriptsize{4.05} & \scriptsize{$2.5\times 10^{13}$} &\scriptsize{$7.40\times 10^{-2}$} \\
\scriptsize{BPL3}&\scriptsize{4.24}  &\scriptsize{$0.5\times 10^{14}$} & \scriptsize{$7.44\times 10^{-2}$}& &  \scriptsize{4.17} & \scriptsize{$0.5\times 10^{14}$} &\scriptsize{$7.44\times 10^{-2}$} \\
 \hline
\end{tabular}
\end{center}

\begin{center}
\scriptsize{\textbf{Table B2. Number of neutrinos above 1 TeV expected per year from the south lobe (see fig. 3  of the left).  Here we show the parameters ($\beta$ and $E_{\nu,b}$)  of neutrino spectrum (eq. \ref{espneu2}) for three broken power laws (BPLs) when one and two UCHERs are expected in PAO.}}\\
\end{center}
\end{document}